\documentclass[prd, onecolumn, nofootinbib, floatfix]{revtex4-1}
\usepackage{silence}
\usepackage{lmodern} 
\usepackage[T1]{fontenc}
\usepackage{textcomp}
\usepackage{xcolor}
\usepackage[scr=boondoxo,frak=boondox,bb=boondox]{mathalfa}
\pdfoutput=1
\usepackage{makecell, multirow}
\setcellgapes{3pt}
\usepackage{hyperref}
\usepackage{amsmath}
\usepackage{tabularx}
\usepackage[vlines]{tabularht}
\usepackage{adjustbox}
\usepackage{tikz}
\usepackage[]{draftfigure}
\usepackage{amssymb}
\usepackage{graphicx}
\usepackage{float} 
\usepackage{color}
\usepackage{color,soul}
\usepackage{ulem}
\usepackage{multirow}
\usepackage{epsfig}
\usepackage{epstopdf}
\usepackage{rotating}
\usepackage{afterpage}
\usepackage{soul}
\usepackage{fancyvrb}
\usepackage{lineno}
\usepackage{amsmath}
\usepackage{xspace}
\usepackage{hyperref}
\usepackage{url}
\usepackage{amssymb,latexsym,mathrsfs}
\usepackage{multirow}
\usepackage{bigints}
\usepackage{soul}
\usepackage{graphicx}
\usepackage{subcaption}
\usepackage{dcolumn}
\usepackage{bm}
\usepackage{epsfig}
\usepackage{caption}
\usepackage{color}
\usepackage{multirow}
\usepackage{makecell}
\usepackage{textcomp}
\usepackage[title]{appendix}
\usepackage{tabularx}

\hypersetup{
    colorlinks=true,
    linkcolor=red,
    citecolor=blue,
} 

\newcommand{\be}{\begin{equation}}
\newcommand{\ee}{\end{equation}}
\newcommand{\bea}{\begin{eqnarray}}
\newcommand{\eea}{\end{eqnarray}}
 
\definecolor{mediumpurple}{rgb}{0.58, 0.44, 0.86}

\definecolor{owngreen}{rgb}{0.0, 0.5, 0.0}

\newcommand{\changes}[1]{\textcolor{black}{#1}}

\newcommand{\Like}{\mathcal{L}}

\begin{document} 
\title{Dark energy reconstruction analysis with artificial neural networks: Application on simulated Supernova Ia data from Rubin Observatory}

\newcommand{\URLSNANA}{\url{https://github.com/RickKessler/SNANA}}
\newcommand{\URLLSST}{\url{www.lsst.org}}
\newcommand{\URLDESC}{\url{https://lsstdesc.org}}
\newcommand{\URLOPSIM}{\url{http://opsim.lsst.org/runs/minion_1016/data/minion_1016_sqlite.db.gz}}
\newcommand{\HOSTLIB}{{\tt HOSTLIB}}
\newcommand{\SALTII}{{\sc SALT-II}}
\newcommand{\DESSN}{DES-SN}
\newcommand{\lowz}{low-$z$~}
\newcommand{\mosfit}{{\tt MOSFiT}\xspace}
\newcommand{\bands}{$ugrizy$}
\newcommand{\PLASTICC}{Photometric LSST Astronomical Time Series Classification Challenge}
\newcommand{\acro}{{\tt PLAsTiCC}}
\newcommand{\SNANA}{{\tt SNANA}}
\newcommand{\pippin}{{\tt Pippin}}
\newcommand{\LSST}{Large Synoptic Survey Telescope}
\newcommand{\DES}{Dark Energy Survey}
\newcommand{\OPSIM}{\verb|OpSim|}
\newcommand{\Spec}{Spectroscopic}
\newcommand{\spec}{spectroscopic}
\newcommand{\specy}{spectroscopically}
\newcommand{\ZPHOT}{{\bf\tt ZPHOT}}
\newcommand{\zSpec}{z_{\rm spec}}
\newcommand{\LCDM}{\Lambda{\rm CDM}}
\newcommand{\OL}{\Omega_{\Lambda}}
\newcommand{\OM}{\Omega_{\rm M}}
\newcommand{\zcmb}{z_{\rm cmb,true}}
\newcommand{\RateUnit}{{\rm yr}^{-1}{\rm Mpc}^{-3}}
\newcommand{\Ftrue}{F_{\rm true}}
\newcommand{\sigF}{\sigma_{F}}
\newcommand{\sigFi}{\sigma_{F,i}}
\newcommand{\sigFtrue}{\sigma_{\rm Ftrue}}
\newcommand{\zSN}{z_{\rm SN}}
\newcommand{\zHOST}{z_{\rm HOST}}
\newcommand{\Ngen}{$N_{\rm gen}$} %

\newcommand{\FiData}{F_i^{\rm data}}
\newcommand{\FiModel}{F_i^{\rm model}}
\newcommand{\xvecSALT}{\vec{x}_5}

\newcommand{\sigFiTilde}{\Tilde{\sigma}_{F,i}}
\newcommand{\logsigmodel}{2\ln({\sigFi/\sigFiTilde})}
\newcommand{\sigzhost}{\sigma_{z,{\rm host}}}

\newcommand{\NSAMPLE}{25} 
\newcommand{\NBIASCORTOT}{$3.1 \times 10^6$}
\newcommand{\NBIASCORHIZ}{$2.6 \times 10^6$}
\newcommand{\NBIASCORLOZ}{$4.4 \times 10^5$}
\newcommand{\NSYST}{7}
\newcommand{\dz}{\delta z}
\newcommand{\fout}{f_{\rm out}}
\newcommand{\sigIQR}{\sigma_{\rm IQR}}
\newcommand{\ztrue}{z_{\rm true}}
\newcommand{\zphot}{z_{\rm phot}}
\newcommand{\zspec}{z_{\rm spec}}
\newcommand{\zcheat}{z_{\rm cheat}}
\newcommand{\zhost}{z_{\rm host}}
\newcommand{\Dz}{\Delta z_{(1+z)}}
\newcommand{\Pfit}{P_{\rm fit}}

\newcommand{\G}{G_{\rm host}}
\newcommand{\mubias}{\Delta\mu_{\rm bias}}
\newcommand{\mutrue}{\mu_{\rm true}}

\newcommand{\dzsyst}{\Delta z_{\rm syst-z}}
\newcommand{\dmusyst}{\Delta\mu_{\rm syst-z}}

\newcommand{\alphaTrueSym}{\alpha_{\rm true}}
\newcommand{\betaTrueSym}{\beta_{\rm true}}
\newcommand{\alphaTrueVal}{0.14}
\newcommand{\betaTrueVal}{3.1}

\newcommand{\sigmu}{\sigma_{\mu}}
\newcommand{\sigmubar}{\overline{\sigma_{\mu}}}

\newcommand{\sigz}{\sigma_{z}}
\newcommand{\sigR}{\sigma_{R}}
\newcommand{\sigint}{\sigma_{\rm int}}

\newcommand{\COVsyst}{{\rm COV}_{\rm syst}}
\newcommand{\COVsysti}{{\rm COV}_{{\rm syst},i}}
\newcommand{\COVstat}{{\rm COV}_{\rm stat}}

\newcommand{\wCDM}{$w$CDM}
\newcommand{\wwCDM}{$w_0w_a$CDM}
\newcommand{\ww}{$w_0$-$w_a$}

\newcommand{\NZBIN}{\textcolor{red}{14}}

\newcommand{\AVGwbias}{\langle w$-bias$\rangle}
\newcommand{\AVGwsigbiaszspecsyst}{$0.025$}
\newcommand{\AVGwsigbiaszphotsyst}{$0.023$}
\newcommand{\AVGwsig}{\langle\sigma_w\rangle}
\newcommand{\STDw}{{\rm STD}_w}

\newcommand{\AVGwwbias}{\langle {w_0}$-bias$\rangle}
\newcommand{\AVGwwsig}{\langle\sigma_{w_0}\rangle}
\newcommand{\STDww}{{\rm STD}_{w_0}}

\newcommand{\AVGwabias}{\langle {w_a}$-bias$\rangle}
\newcommand{\AVGwasig}{\langle\sigma_{w_a}\rangle}
\newcommand{\STDwa}{{\rm STD}_{w_a}}

\newcommand{\AVGFoM}{{\langle}{\rm FoM}{\rangle}}

\newcommand{\RatioFoM}{{\cal R}_{{\rm FoM},i}}
\newcommand{\FoM}{{\rm FoM}}
\newcommand{\FoMStat}{{\rm FoM}_{\rm stat}}
\newcommand{\FoMSysti}{{\rm FoM}_{{\rm syst},i}}
\newcommand{\AVGFOMzspecstat}{$136$}
\newcommand{\AVGFOMzspecsyst}{$95$}
\newcommand{\AVGFOMzphotstat}{$237$}
\newcommand{\AVGFOMzphotsyst}{$145$}
\newcommand{\AVGwbiaszphotsyst}{$0.0091$}
\newcommand{\AVGwabiaszphotsyst}{$0.0363$}
\newcommand{\AVGwbiaszspecsyst}{$0.0140$}
\newcommand{\AVGwabiaszspecsyst}{$0.0662$}

\author{Ayan Mitra${}^{1,2,3}$} 
\email{ayan@illinois.edu}
\author{Isidro G\'omez-Vargas${}^{4,5}$ } 
\email{isidro.gomezvargas@unige.ch}
\author{Vasilios Zarikas${}^{6}$ } 
\email{vzarikas@uth.gr}

\affiliation{
${}^1$Center for AstroPhysical Surveys, National Center for Supercomputing Applications, University of Illinois Urbana-Champaign, Urbana, IL, 61801, USA\\
${}^2$Department of Astronomy, University of Illinois at Urbana-Champaign, Urbana, IL 61801, USA\\
${}^3$Kazakh-British Technical University, 59 Tole Bi Street, 050000 Almaty, Kazakhstan\\
${}^4$Instituto de Ciencias F\'isicas, Universidad Nacional Aut\'onoma de M\'exico, 62210, Cuernavaca, Morelos, M\'exico. \\
${}^5$Department of Astronomy of the University of Geneva, 51 Chemin Pegasi, 1290 Versoix, Switzerland.\\
${}^6$Department of Mathematics, University of Thessaly, Lamia 35132, Greece
}

\date{\today} 

\begin{abstract}
\changes{In this paper, we present an analysis of Supernova Ia (SNIa) distance moduli ($\mu(z)$) and dark energy using an Artificial Neural Network (ANN) reconstruction based on LSST simulated three-year SNIa data. The ANNs employed in this study utilize genetic algorithms for hyperparameter tuning and Monte Carlo Dropout for predictions. Our ANN reconstruction architecture is capable of modeling both the distance moduli and their associated statistical errors given redshift values. We compare the performance of the ANN-based reconstruction with two theoretical dark energy models: $\Lambda$CDM and Chevallier-Linder-Polarski (CPL). Bayesian analysis is conducted for these theoretical models using the LSST simulations and compared with observations from Pantheon and Pantheon+ SNIa real data. We demonstrate that our model-independent ANN reconstruction is consistent with both theoretical models. Performance metrics and statistical tests reveal that the ANN produces distance modulus estimates that align well with the LSST dataset and exhibit only minor discrepancies with $\Lambda$CDM and CPL.
}
\end{abstract}

\maketitle
\section{Introduction}
\label{sec:intro}

The discovery of the accelerated expansion of our Universe using Supernova Ia observations revolutionized our understanding of modern cosmology \citep{riess_1998, Perlmutter:1998np, 2008ApJ...686..749K, Betoule2014, Planck2016, 2012ApJ...746...85S}. The standard model of cosmology, $\Lambda$CDM, is composed of ordinary matter, a dark energy modeled as a cosmological constant $\Lambda$ that is responsible for the acceleration of the Universe, and Cold Dark Matter (CDM) \cite{Carroll:2000fy} that shapes the cosmic structure through gravitational influence. $\Lambda$CDM  assumes a homogeneous and isotropic Universe and has been in excellent agreement with most of the currently available data. However, it has its theoretical drawbacks, such as fine-tuning and cosmic coincidence \cite{SolaPeracaula:2022hpd}, 
 \changes{and observationally  it suffers from the Hubble tensions between the values of $H_0$  obtained from CMB observations \cite{Planck2018} and SNIa observations \cite{pantheon_new, Pantheon, Riess_2022, DiValentino:2021izs, Aluri:2022hzs, Perivolaropoulos:2021jda, Kamionkowski:2022pkx, DiValentino:2021izs, Krishnan:2021dyb, Abdalla:2022yfr, Krishnan:2021jmh, Smith:2022hwi}}. These issues open the door to study models beyond the standard cosmological model.     %
    
There are non-parametric inference techniques that try to avoid assumptions of specific theoretical models, to find new dark energy properties from the data. They extract information directly from the data and infer unknown quantities based mainly on the data with as few assumptions as possible \cite{sahni2002cosmological, wasserman2006all}. Some interesting examples of these types of methods previously used to reconstruct cosmological functions are Principal Component Analysis\cite{sharma2020reconstruction}, smoothed step functions \cite{r8}, Gaussian processes \cite{williams2006gaussian, Keeley:2020aym, l2020defying, r10, escamilla2023model} and extrapolation methods \cite{montiel2014nonparametric}. These approaches are also known as model-independent reconstructions, and they can be considered as new models based on the data, they can be used to analyze their similarity with different theoretical models and, to determine what model describes better the observational data.
There are several works based on model-independent approaches to analyze dark energy features \cite{sahni2006reconstructing, holsclaw2010nonparametric, zhao2017dynamical, r8}, cosmic expansion \cite{montiel2014nonparametric}, deceleration parameter \cite{r10}, the growth rate of structure formation \cite{l2020defying, gomez2023neuralrecs} and luminosity distance \cite{wei2017improved, lin2019non, wang2020reconstructing}. \changes{In particular, Dark Energy has been studied using neural networks \cite{gomez2023neuralrecs, escamilla2020deep, wang2020reconstructing, dialektopoulos2023neural}, genetic algorithms \cite{arjona2020machine, arjona2020can, arjona2020hints, arjona2022testing}, and other machine learning techniques \cite{escamilla2023model, Keeley:2020aym}.}

In current times, the increase in computing power, bigger telescopes, enhanced CCDs, and a large amount of observational data have allowed the incorporation of machine learning methods as analysis tools in observational cosmology \cite{lin2017does, peel2019distinguishing, arjona2020can, wang2020machine, gomez2023neuralgenetic, chacon2023analysis, medel2023cosmological}. Artificial Neural Networks (ANN) are one of the most revolutionary methods of Artificial Intelligence (AI), and they have been successfully employed in a wide range of applications in cosmology, for example, in image analysis \cite{dieleman2015rotation, ntampaka2019deep}, N-body simulations \cite{chacon2023analysis, rodriguez2018fast, he2019learning} and statistical methods \cite{gomez2023neuralrecs, gomez2023neuralgenetic, alsing2019fast, li2019model, hortua2020constraining, hortua2020parameter}.  

\changes{We are using neural networks to generate model-independent models based on the data with a methodology presented in previous works \cite{gomez2023neuralrecs, gomez2023neuralgenetic}. A key advantage of well-trained neural networks is their independence from a fiducial cosmology, allowing the generated data to be treated as new observations reflecting the true nature of the original dataset. Furthermore, neural networks surpass standard interpolation techniques by not requiring any statistical distribution assumptions for the data due to their nonlinear modeling capabilities.}
From the data perspective, the forthcoming observational programs are anticipated to enhance the dark energy precision by almost ten folds \cite{FoM}, some examples are Euclid \cite{EuclidTheoryWorkingGroup:2012gxx,https://doi.org/10.48550/arxiv.0912.0914}, LSST \cite{LSSTDarkEnergyScience:2018jkl, Zhan:2017uwu, LSST:2008ijt}, Roman Space Telescope \cite{Akeson:2019biv}, the Thirty Meter Telescope (TMT) \cite{TMT}, and the already operational JWST \cite{Gardner_2006}.
Consequently, all the reasons above culminate in the increasing effort of leveraging AI and large datasets to investigate dark energy in a new way. This approach aims to overcome the longstanding issues of model degeneracy and the tension between different cosmological measurements without relying on predetermined theoretical models.

In this paper, we present a dark energy reconstruction analysis based on the state-of-the-art LSST's three years of simulated \acro \ SNIa data \cite{plasticc_announce} and Artificial Neural Network modeling. We generated \changes{data-based models with neural networks} for the distance modulus and we compared this model with the Bayesian parameter estimation of dark energy models. In addition, we contrast the results with the SNIa data from the Pantheon and Pantheon+ surveys as real observational data. Our Bayesian analysis allows us to explore whether a model preference can be inferred from the LSST dataset, between the two most prevalent formulations of dark energy: $\Lambda$CDM and CPL (see \S\ref{sec:demodels}). 

This paper is structured as follows, in sec.~\ref{sec:background} we present the theory of the two different dark energy models we adopted in this analysis and a brief overview of the ANNs. In  sec.~\ref{sec:lsst} we outline the LSST survey program, which leads us to the next section \ref{sec:datasets}, where we introduce the LSST simulated SNIa dataset; which is the primary dataset for our analysis. We also introduce the additional Pantheon SNIa datasets later in this section, used for comparisons only (and no ANN training is done with them). In the next sec.~\ref{sec:method}, we discuss the methods used for constructing the ANN network and how the model is trained. In sec.~\ref{sec:results} we present the findings of our analysis and finally sec.~\ref{sec:conclude} presents the conclusion of this paper. 

\section{Background}\label{sec:background}


\subsection{Dark Energy Models}
\label{sec:demodels}

We have considered the two most popular cosmological dark energy models, the $\Lambda$CDM and the Chevallier-Linder-Polarski (CPL) model \citep{eos1, eos2}, to test our LSST neural reconstruction in this analysis. The $\Lambda$CDM model often also called the concordance model, assumes that the Universe's energy density is driven by a cosmological constant $\Lambda$ and it has a dark matter characterized as Cold Dark Matter (CDM). It is this cosmological constant that is believed to be due to dark energy. The dark energy component is parameterized by its Equation of State (EoS) parameter $w$ and is defined as:
\bea
w = \frac{p}{\rho},
\eea
and a cosmological constant $\Lambda$ corresponds to $w=-1$ and $\rho$ (energy density) constant. Over the last two decades, the $\Lambda$\text{CDM} model has consistently been shown to have good agreement with the observations. \changes{For a flat Universe,} the Hubble parameter relates to the redshift ($z$) as,
\begin{equation}
H^2(z)=H_0^2\left[\Omega_{m}(1+z)^3+(1-\Omega_{m})\right]. 
\end{equation}
where $\Omega_m$ is the matter density. 

The second dark energy model we used in this analysis, is the CPL model. Parameterized as a function of scale factor $a$ \footnote{Scale factor $a$ and redshift $z$ is linked as $a=(1+z)^{-1}$.}, it is a two-parameter dark energy model. 
\bea
w = w_0 + w_a(1-a),
\eea
it can be seen as a first-order approximation of the  Taylor expansion of a more general dark energy EoS  \(w(a)\) \citep{2017ApJ...850..183Z}. This approach ensures that the model remains well-behaved across the entire range of the Universe's history, from high to low redshifts. The dynamics of the Universe's expansion are encapsulated by the scale factor \(a\), with the background evolution described as \changes{(for a flat Universe)},
\begin{equation}
H^2(a) = H_0^2   
\left[ \frac{\Omega_{m0} + a^3(1 - \Omega_{m0})a^{-(3w_0 + 3w_a + 3)}\exp\left(3w_a(a - 1)\right)}{a^3} \right].
\end{equation}

The CPL is very well-behaved, all the way, from $a=0$ to $a=1$. Being a simple two-parameter phenomenological description of dark energy,  the CPL is easy to analyze and also it adapts well to a wide variety of dark energy models.

\subsection{Supernova Ia}
\label{sec:snia}
Type Ia supernovae (SN Ia) are believed to be the result of the thermonuclear disruption of
 carbon-oxygen white dwarfs which reaches the Chandrasekhar-mass limit of stability ($M_{Ch} \sim 1.4$M solar mass) by accreting matter from a companion \cite{1960ApJ...132..565H}. SNIa light curves show remarkable homogeneity after correcting for stretch \changes{($x_1$, i.e. the time stretching of the light curve) and SN color ($c$) at the peak brightness parameters.} Observed flux from SNIa can be used to compute the luminosity distance $D_L$. For an FLRW Universe, with a dark energy component EoS $w(z)$ the $D_L$ is given as,  
\begin{equation}
D_L =  (1+z) \frac{c}{H_0} \int_{0}^{z} \frac{dz'}{E(z')}
\end{equation}
where distance $D_L$ is in megaparsecs (Mpc) and $E(z')$ is given as,
\begin{equation}
    E(z) = \sqrt{\Omega_m (1+z)^3 + (1 - \Omega_m) \exp\left(3 \int_{0}^{z} \frac{1 + w(z')}{1 + z'} dz'\right)}
\end{equation}

Distance estimations in SNIa are based on an empirical observations, that all these events belong to a homogeneous class, whose variablilities can be captured mostly via the stretch and the color component. 
\changes{
For each event, the measured distance modulus ($\mu$) is based on the Tripp formula \cite{Trip1998},
\begin{equation}
   \mu = m_B + \alpha x_1 - \beta c  + M_0 + \mubias~,   \label{eq:mu}
\end{equation}
where $m_B$ ($= -2.5\log_{10}(x_0)$) is the approximate B-band peak brightness magnitude of the SN. $m_B$ along with $x_1$ and $c$ are the SALT3 light curve parameters. While $\alpha$ and $\beta$ are global nuisance parameters which sets the amplitude of the stretch-luminosity and color-luminosity corrections respectively. $M_0$ is the absolute magnitude of the SNIa, with $x_1=0$, $c=0$, and is also a global nuisance parameter.  Without a calibrated absolute distance scale, $M_0$ is degenerate with the Hubble parameter $H_0$.   The final term  $\mubias=(\mu-\mu_{true})$,  accounts for the corrections resulting from selection effects and analysis biases and is determined from a separate biasCor simulation in a 5 dimensional space of $\{z,x_1,c,\alpha,\beta\}$ \cite{bbc}. }
 
Over the last two decades, numerous studies have utilized Type Ia Supernovae (SNIa) as the primary probe for investigating dark energy \cite{panstarrs,Betoule2014,s1,Dhawan:2021ztf}. 
 Through successive SNIa programs, our precision in understanding dark energy has significantly improved, with efforts still ongoing \cite{Kessler2010, Linder2019,Mitra2021}.

\subsection{Dark Energy Reconstruction and Artificial Neural Networks}
\label{sec:anns}

 Since the renewed interest in dark energy study, post the discovery of accelerating Universe via the SNIa observations \citep{riess_1998, Perlmutter:1998np}, the effort to reconstruct dark energy has been a very widely cultured topic \cite{sahni2006reconstructing,re1,re2,re3,re4,re5,re6,re7}. Broadly, the reconstruction process can be classified either under parametric or non-parametric category. The parametric form of dark energy reconstruction has seen a lot of success, and it provides a lot of advantages, such as simplicity and interpretability, consistency in predictive power, especially in extrapolating beyond observed data, their simple forms allow easy efficiency of parametric models in fitting data. However parametric models also come with disadvantages. The biggest drawback of parametric models is the critical issues of model dependence and potential biases \cite{Liddle2007}. They also provide limited flexibility, in capturing unexpected features of the dark energy equation of state \cite{Copeland2006}, and finally, additional complexities can arise from parameter degeneracies, leading to over or underfitting \cite{Trotta2008}. 

While parametric reconstruction methods have provided valuable insights by imposing specific models on the dark energy EoS $w(z)$, they inherently assume a certain level of theoretical bias toward the functional form of dark energy as discussed above. These limitations fanned the study of non-parametric reconstruction methods, which offer a more model-independent avenue for analyzing dark energy. Unlike their parametric counterparts, non-parametric methods do not rely on predefined equations or functions to describe the reconstructed parameter \cite{r8,r9,r10,r11,r12}. Instead, they leverage the data directly, allowing the underlying properties to manifest more freely in the analysis. In context to dark energy, this shift towards non-parametric reconstruction approaches can significantly open up new possibilities by reducing the theoretical prejudices and using the complexity and richness of the observational data \cite{r13,r14,gomez2023neuralgenetic}. In this analysis, we use ANN-based non-parametric reconstruction technique \cite{gomez2023neuralrecs, Wang:2020sxl, Wang:2019vxv} to obtain distance modulus ($\mu$) from SNIa observations. 
ANNs are powerful computational models capable of approximating any continuous nonlinear function, as demonstrated by \cite{hornik1990universal}, making them exceptionally suited for modeling complex and large datasets. They are also better than the Gaussian Process (GP) based reconstructions in the sense that the ANN approach offers enhanced flexibility (eg. overfitting issues, kernel selection sensitivity, etc.) and accuracy in modeling natural processes compared to GP by having more optimizable hyperparameters. They are also less influenced by any prior assumptions which GP is more sensitive of \cite{dialektopoulos2022neural}.  
  
For a comprehensive background on neural networks, seminal texts such as \cite{goodfellow2016deep, bishop2006pattern, nielsen2015neural} provide in-depth analyses. The details of our ANN-based architecture are outlined in sec.\S\ref{sec:method}.

\section{LSST: Overview}\label{sec:lsst}
The Legacy Survey of Space and Time (LSST), conducted by the Vera C. Rubin Observatory collaboration, represents a paradigm shift in astrophysical surveys with its unparalleled scope and technical sophistication. Scheduled to begin in \changes{2025}, the LSST will deploy an advanced observational apparatus, featuring an $8.4$ m primary mirror with a $6.7$ m effective aperture and a state-of-the-art $3200$-megapixel camera, yielding a wide field of view of $9.6$ square degrees. The LSST is designed to survey approximately 18,000 square degrees of the southern sky over a decade, utilizing six optical passband filters to facilitate deep, wide, and fast observations. The survey aims to amass over 32 trillion observations of 20 billion galaxies and a similar number of stars, achieving a depth of $24^{th}$ magnitude in its six filter bands, spanning wavelengths from ultraviolet to near-infrared \cite{Cahn2009, ivezic}. These efforts are projected to catalog millions of supernovae, among other transient phenomena, offering an unprecedented dataset for probing the dark Universe. The Dark Energy Science Collaboration (DESC) \footnote{\url{https://lsstdesc.org/}}, comprising nearly 1,000 members, intends to harness this vast trove of data to extract high-precision measurements of fundamental cosmological parameters, leveraging prior data challenges to refine analysis pipelines in anticipation of the survey's extensive data output \citep{dc2, Sanchez}. 

\section{Datasets}\label{sec:datasets}
Our reconstruction analysis using the ANN models is based on the LSST \changes{like} a 3-year simulated SNIa data. We have, also for comparison, analyzed the performance of the considered dark energy models using Pantheon and Pantheon+ SNIa datasets, all of which are described in this section below. 

\subsection{Simulated LSST SNIa Data}\label{sec:sim}
We use three years of simulated Type-Ia Supernovae data from the LSST. This data is derived from \cite{mitra2023using} and it consists of 5785 data points of SNIe with their simulated covariance matrix of statistical and systematic errors combined, within the redshifts $0.01<z<1.4$.

The SN data is generated using the LSST DESC time domain (TD) pipeline and \SNANA\ code \citep{snana}, consisting of four main stages  (illustrated in Fig. \changes{4} of \cite{mitra2023using}). These include SN brightness standardization via a Light Curve (LC) fit stage, simulations for bias correction, and a BBC stage for Hubble diagram production before the last stage, cosmology fitting. We used a  Hubble diagram and the associated covariance matrix (statistical + systematic) produced from the BBC stage to perform the cosmological fitting. 

For mock generation, input cosmology: $\Omega_m=0.3150, \, \Omega_\Lambda=0.6850, \, w_0=-1, \, w_a=0$ was used. The curvature is computed internally as $\Omega_k=1-\Omega_m-\Omega_\Lambda$ \changes{(for our analysis case this is consistent with the assumptions of a flat cosmology)}. It is based on cosmological parameters from Planck 2018 \cite{Planck2018}. In addition, the parameter $H_0$ is set to $70.0$ km/s/Mpc, this value is tied to SALT2 training \cite{salt} and we use the SALT2 lightcurve model\footnote{Like mentioned in sec.~\ref{sec:snia} that $H_0$ has a degeneracy with $M_0$ and hence the choice of $H_0$ value is not significant.}. 


We use a Hubble diagram (HD) \changes{containing redshift, distance modulus information} alongside its corresponding covariance matrix.  The datasets are composed of a mixture of spectroscopically ($\zspec$) and photometrically ($\zphot$) identified SNIa candidates at low redshift and high redshift respectively. 

The photometric redshift determination and its uncertainty were based on \cite{Graham2018_photoz} while using host galaxy photo-$z$ as priors (\cite{Kessler2010}). \cite{mitra2023using} re-simulated the \acro \  data based on the DDF strategy of the LSST \footnote{\changes{\url{https://www.lsst.org/scientists/survey-design/ddf}}}, augmented with low-redshift spectroscopic data from the DC2 analysis \changes{\cite{Sanchez}}. \changes{Based on \acro\ ~\cite{plasticc_announce} ``$\zspec$'' sample is composed of two sets of events whose spectroscopic redshifts have an accuracy of $\sigma_{z} \sim 10^{-5}$. The first subset is made up of \specy\ confirmed events whose redshift is predicted to be accurate by the 4MOST spectrograph ~\cite{4MOST2}, which is being built by the European Southern Observatory and is expected to become operational in $2023$. 4MOST is situated at a latitude similar to that of the Rubin Observatory in Chile. The second subset is composed of photometrically identified events with an accurate host galaxy redshift determined by 4MOST. The second subset has about $\sim60$\% more events than the first subset.
For the photometric sample, host galaxy photo-$z$  was used as a prior (adapted from ~\cite{Kessler2010}).  The photometric redshift and rms uncertainty was based on ~\cite{Graham2018_photoz}. The whole simulation was re-done based on the \acro \  DDF data\footnote{The LSST has different observing strategies, the deep field and the wide field, called as the DDF (Deep drilling field) and the WFD (Wide Fast Deep) respectively.}. Additional low redshift spectroscopic data was used from the DC2 analysis (simulated with WFD cadence \footnote{While WFD cadence is being used for the DC2 analysis, a full scale WFD study is yet not done, and is underway currently in the LSST.}).}
The study by \cite{mitra2023using}, which simulated exclusively SNIa without considering contamination, generated a statistical plus systematic covariance matrix to account for seven systematics. The HD is comprised of 5785 SNIa candidates. Importantly, the work by \cite{mitra2023using} expanded on the initial analysis plan for the LSST outlined by \cite{LSSTDarkEnergyScience:2018jkl}, which focused on a 1 and 10-year timeline for studying Supernova Type Ia (SNIa) cosmology.  
\changes{Since this study is based on three years simulated SNIa from the LSST using the DDF observations only hence the smaller sample size, as opposed to millions that is forecast to observe during its ten years study using both the DDF and the Wide-Fast-Deep (WFD) observations combined.} 
\cite{mitra2023using} did this by conducting a comprehensive analysis that takes into account both the spectroscopic and photometric redshifts of the supernova's host galaxies.

\subsection{Pantheon and Pantheon+ compilations}

The Pantheon datasets are all publicly available in the form of unbinned Hubble diagram data. 
Pantheon has 1048 data points \cite{scolnic2018complete} within redshifts between $z=0.01$ and $z=2.3$, and Pantheon+ has 1550 Type Ia  between $z=0.001$ and $z=2.26$ \cite{brout2022pantheon+}. They are the latest compilations of the SNIa observations, expanding from the previously concluded confirmed Pan-STARRS1 survey \cite{2002SPIE.4836..154K}. Unlike the \changes{5785 LSST simulated SNeIa}, all of Pantheon's data are spectroscopic in nature \changes{and it has slight higher redshift range than the redshift range that the LSST is forecast to observe.} 

\section{Methodology}\label{sec:method}
\begin{figure}[!ht]
     \centering
     \includegraphics[width=0.4\textwidth]{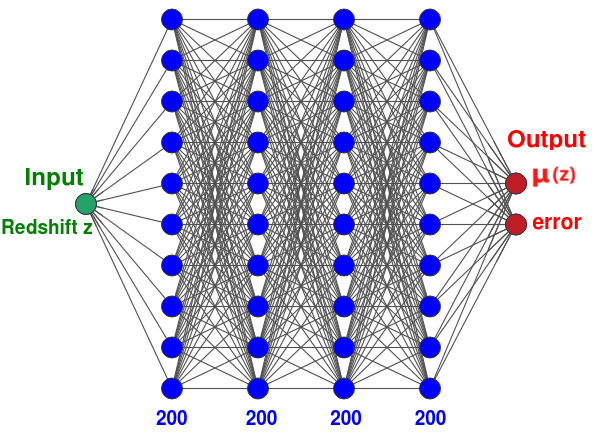}
     \caption{ANN architecture founded by the hyperparameter tuning with genetic algorithms using \texttt{nnogada}.}
     \label{fig:ann_arch}
 \end{figure}

This section describes how we use the ANN to generate neural network reconstructions for the distance modulus. We employ the methodology presented in \cite{gomez2023neuralrecs} using the hyperparameter tuning suggested in \cite{gomez2023neuralgenetic}. \changes{We normalize the dataset, and split the 5785 LSST simulated SNeIa in 80\% for training and 20\% for validation to monitor possible under or overfitting. The loss function used is Mean Squared Error (MSE) and we train our neural networks through 2000 epochs.} In our case, we use the \texttt{TensorFlow} library to implement our neural network model. Below we outline the steps:

\begin{itemize}
    \item \changes{We perform an exploratory analysis of the dataset and then prepare the dataset for training our neural network.} We consider the redshift ($z$) as input, and the distance modulus ($\mu$) with the error ($\sigma(\mu)$) conformed by the sum of their statistical and systematic errors.
    
    \item We train different architectures of neural networks with Monte Carlo Dropout \cite{gal2015dropout} (as in \cite{gomez2023neuralrecs}) to gain empirical knowledge about the range of hyperparameters. \changes{The Monte Carlo Dropout method randomly deactivates neurons during predictions after the neural network has been trained. This approach allows for estimating the uncertainties in the neural network's predictions, is mathematically equivalent to approximating variational inference in deep Gaussian processes, and helps create more robust models by effectively reducing the risk of overfitting.} We use the Monte Carlo dropout implementation from AstroANN \cite{astronnleung2019deep, astronn}. \changes{The use of Monte Carlo Dropout introduces probabilistic behavior in neural networks and aligns with the Bayesian neural networks paradigm \cite{gal2015dropout, benatan2023enhancing}. Predictions generated using this method should be represented by the mean and standard deviation of multiple predictions, each with different neurons turned off.}

    \item To find a proper combination of hyperparameters, we set the search space for the use of \texttt{nnogada} \citep{gomez2023neuralgenetic} and their implemented canonical genetic algorithms with elitism from the DEAP library \cite{DEAP_JMLR2012,de2014deap} (for details on genetic algorithms we recommend Ref. \cite{medel2023cosmological}). In our case, we consolidated the search space with the number of layers $\in [3, 4]$, the number of neurons for hidden layers $\in [100, 200]$, and batch size $\in [8, 16, 32, 64]$. 
    
    \item The best combination found was 4 layers with 200 neurons in each one and a batch size of 32 with \changes{the MSE loss function value of 0.02528} (see Fig.~\ref{fig:ann_arch}).
    
    \item \changes{We use the best configuration to build our artificial neural network and train it with LSST simulated data.}
    
    \item Once the neural network is trained, we perform several predictions of the distance modulus \changes{using different redshift values, within the range of the training data,} to generate a non-parametric reconstruction based on the LSST simulated SNIa data.
    
    \item \changes{To have an insight into the cosmological models considered in this work constrained by the LSST data, we perform an initial Bayesian inference using nested sampling with the \texttt{dynesty} library \cite{speagle2020dynesty} and the \texttt{SimpleMC} code \cite{aubourg2015, simplemc}. We also do it with Pantheon and Pantheon+, to compare the results obtained with the LSST simulations. This provides us with a baseline reference, before the subsequent ANN analysis.} We compare the theoretical predictions and the reconstruction with the data, and we measure with the MSE metric to test which cases are in more agreement with the dataset. The values of the dark energy model parameters are previously fitted with Bayesian inference using the same data used for the neural reconstruction.
\end{itemize}

\changes{Given that the available LSST data is simulated and, for simplicity, we are using statistical errors of the distance modulus with a diagonal covariance matrix, we acknowledge that approaches such as those in \cite{gomez2023neuralrecs} consider modeling non-diagonal covariance matrices; however, this is beyond the scope of our paper. During training, the ANN learns to predict the distance modulus along with its statistical errors given a redshift value from the dataset. Consequently, predictions are provided for each redshift corresponding to the pair (distance modulus, error). To generate a point in the reconstruction, we employ Monte Carlo Dropout, which uses the trained neural network to make multiple predictions by randomly deactivating different neurons. We then calculate the mean and standard deviation of these predictions, with the latter representing the intrinsic error from the ANN. This intrinsic error is added to the modeled statistical error.}

\changes{We use the mean squared error (MSE) to compare the similarities between the neural reconstruction predictions, the theoretical models, and the SNIa compilations, as it offers a straightforward method for this analysis. Although other approaches, such as calculating Bayesian evidence or information criteria, are available for model comparison, we do not employ them due to the complexity and large number of parameters in neural networks. Our focus is on determining which theoretical model is most similar to the neural reconstructions, rather than seeking to develop a model superior to the theoretical ones.}

For interested readers, the GitHub repository\footnote{\url{https://github.com/igomezv/LSST\_DE\_neural_reconstruction}} is available, with the data and the Python notebooks in which the hyperparameter tuning and reconstructions were performed.

\section{Results \& Discussions}\label{sec:results} 

\subsection{Bayesian Analysis}
\begin{figure*}[htbp]
    \centering
    \begin{subfigure}[b]{0.42\textwidth}
        \centering
        \includegraphics[width=\textwidth]{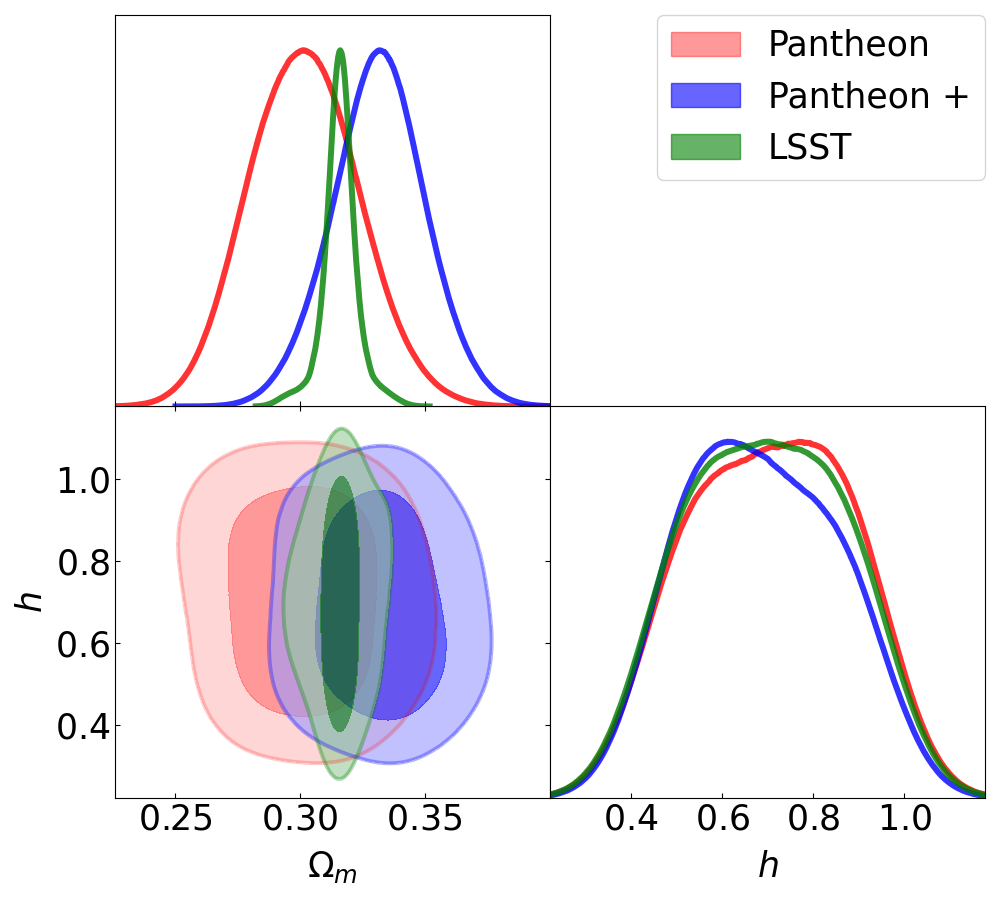}
        \caption{$\Lambda$CDM}
        \label{fig:cornerLCDM}
    \end{subfigure}
    \hfill
    \begin{subfigure}[b]{0.55\textwidth}
        \centering
        \includegraphics[width=\textwidth]{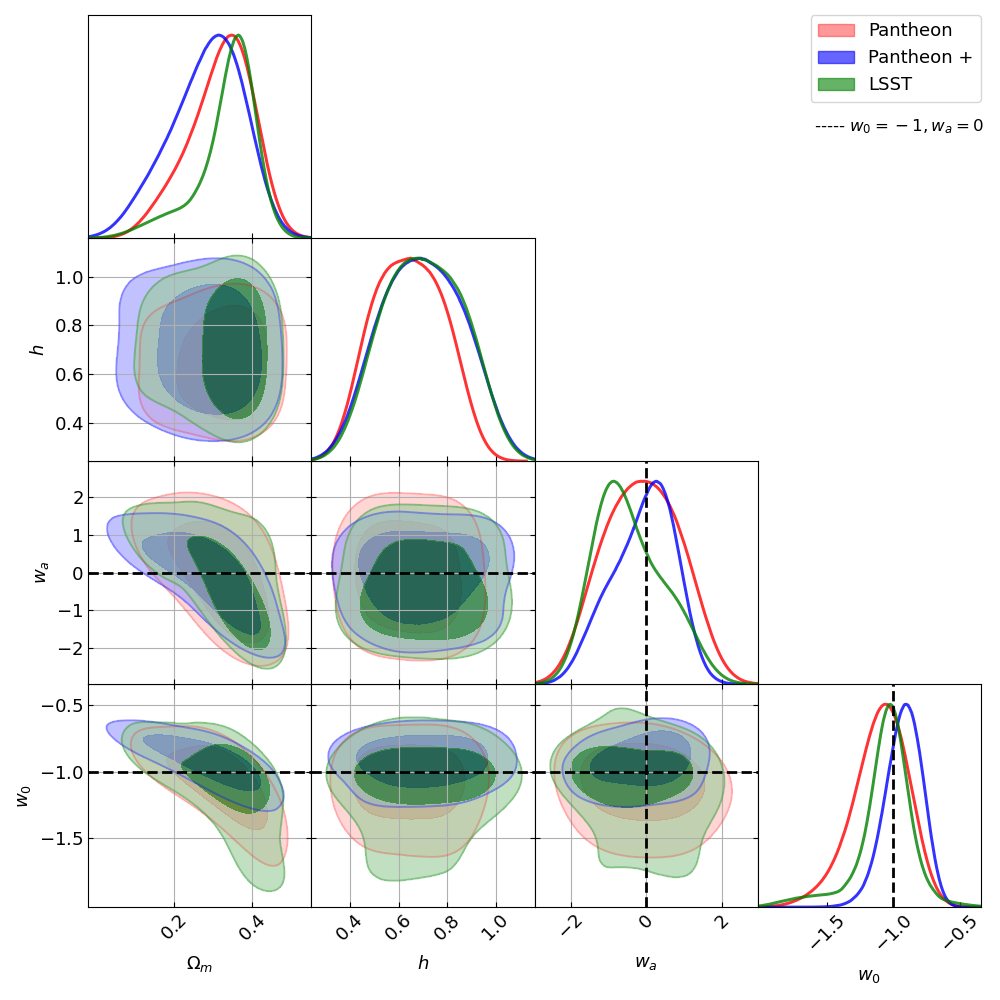}
        \caption{CPL}
        \label{fig:cornerCPL}
    \end{subfigure}
    \caption{One and two-dimensional posterior distributions from Bayesian sampling for the free parameters of
$\Lambda$CDM (left) and CPL (right) using SNIa data from Pantheon \changes{(1048 SNIa)}, Pantheon+ \changes{(1550 SNIa)}, and LSST simulations \changes{(5785 SNIa)}. \changes{For reference, dashed lines indicate the theoretical $\Lambda$CDM values of $w_0 = -1$ and $w_a = 0$.}}
    \label{fig:cornerplots}
\end{figure*}

\changes{As we mentioned in Section \ref{sec:method}, we perform a Bayesian parameter estimation of the $\Lambda$CDM (constrained to be flat) and CPL models using our LSST simulated data, Pantheon and Pantheon+ surveys.} Table \ref{tab:bayesian} includes the ranges of the priors used for the free parameters of both dark energy models. In addition, it shows the results of posterior distribution sampling and the Bayes factor to compare models. According to Fig.~\ref{fig:cornerplots} and Table \ref{tab:bayesian}, we can notice that for $\Lambda$CDM the $\Omega_m$ is considerably better constrained with the LSST data than the other two SNIa compilations; while the $h$ parameter is practically not affected between the three different datasets. On the other hand, for the CPL model, the parameter estimation with LSST data is consistent with Pantheon and Pantheon+, with a better constraint for the $w_0$ parameter. 

\begin{figure*}[t!]
    \centering
    \begin{subfigure}[b]{0.3\textwidth}
        \centering
        \includegraphics[width=\textwidth]{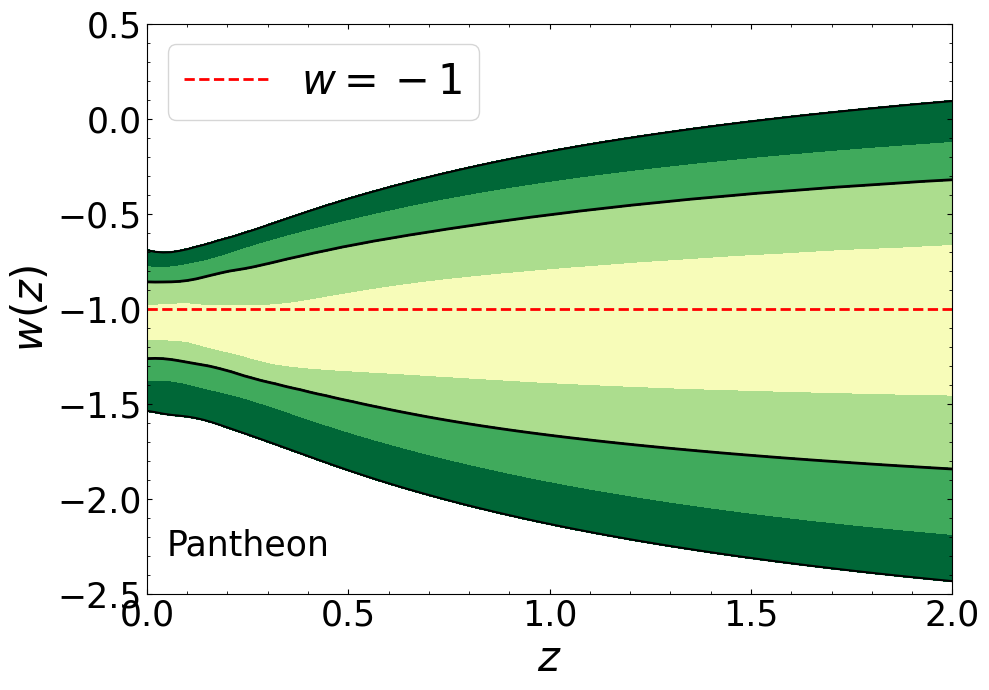}
        \caption{Pantheon}
        \label{fig:Pl}
    \end{subfigure}
    \hfill
    \begin{subfigure}[b]{0.3\textwidth}
        \centering
        \includegraphics[width=\textwidth]{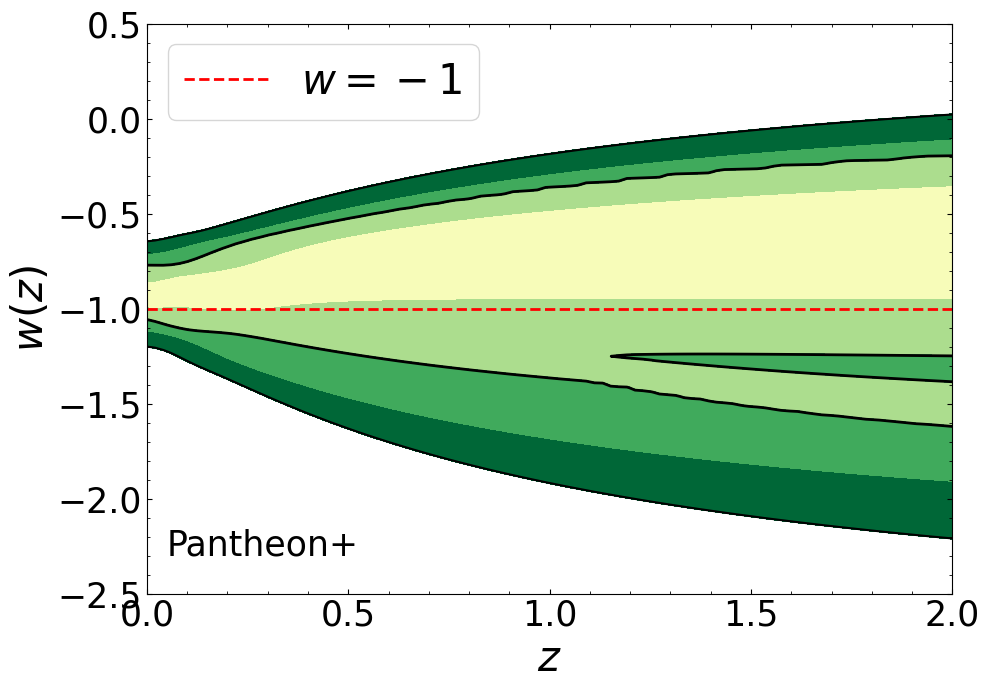}
        \caption{Pantheon+}
        \label{fig:image2}
    \end{subfigure}
    \hfill
    \begin{subfigure}[b]{0.3\textwidth}
        \centering
        \includegraphics[width=\textwidth]{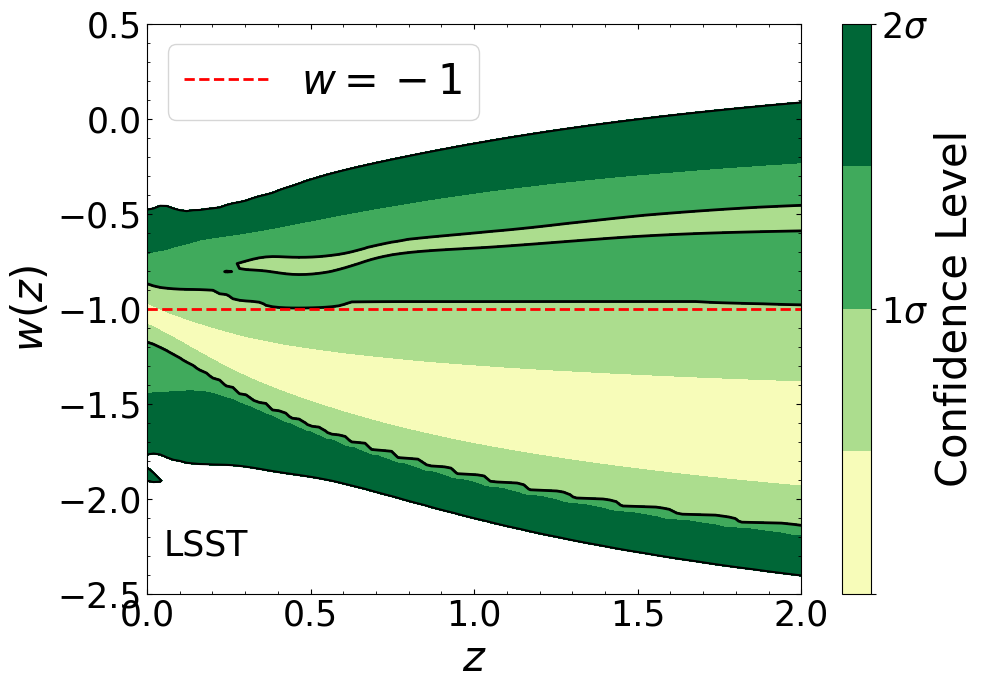}
        \caption{LSST}
        \label{fig:image3}
    \end{subfigure}
    \caption{EoS for CPL with the three different datasets. We obtained these plots from the posterior distribution sampling using \texttt{fgivenx} \cite{handleyfgivenx}. It is interesting to note that for the LSST case, the $w=-1$ line (red dashed) marginally crosses the $1-\sigma$ range.}
    \label{fig:fgivenx}
\end{figure*}

\changes{In all the cases the $-2\log \Like_{max}$, equivalent to the chi-squared,  (Table~\ref{tab:bayesian}) obtains a lower value for the LSST dataset, which implies that this data obtains a tiny better fit for the cosmological models}. In addition, something interesting is that using the LSST data, the Bayesian model comparison suggests that $\Lambda$CDM increases its advantage over CPL because according to Jeffrey's scale, both Pantheon and Pantheon+ $\Lambda$CDM have \changes{moderate evidence with Bayes factors around 1.5, and for LSST data this scale is also moderate, but with a Bayes factor around 2.7, close to the boundary between moderate and strong. Please consider that we are using natural logarithm for Likelihood, Bayesian evidence, and Bayesian factor}. This behavior can also be analyzed in Fig. \ref{fig:fgivenx}, where the EoS for CPL using LSST data puts the $w=-1$ line for $\Lambda$CDM almost out of $1-\sigma$, while for Pantheon and Pantheon+, the most probable value for $w$ is statistically closer to the $\Lambda$CDM prediction. However, considering only the values of $-2\log \Like_{max}$, for the three datasets, CPL has always a better fit to the data than $\Lambda$CDM; therefore, the Bayesian evidence could be very sensitive to the extra-parameters of CPL and, in future work, it could be worth to incorporate other model comparison techniques because Bayesian evidence has received some criticisms in cosmological data analysis \cite{efstathiou2008limitations, nesseris2004comparison, koo2022bayesian}. However, our results clearly show that the LSST data can give more advantages while investigating for model selection tests in the future (more so with LSST $10$ years  SNIa data) though in the current analysis our results from Table~\ref{tab:bayesian}, \changes{are insufficient to decide in favor of any model}.  

\begin{table*}[t!]
\captionsetup{justification=raggedright, singlelinecheck=false, font=footnotesize}
    \centering
    \begin{tabular}{|c|c|c|c|c|c|c|c|}
    \hline
                    & Priors  & $\Lambda$CDM Pantheon & CPL Pantheon & $\Lambda$CDM Pantheon+  & CPL Pantheon + & $\Lambda$CDM LSST & CPL LSST \\
         \hline
         $\Omega_m$ & [0.05, 0.5] & $0.3017 \pm 0.0220$ & $0.3203 \pm 0.0821$ & $0.3317 \pm 0.0184$ & $0.2846 \pm 0.0930$ & $0.3163 \pm 0.0049$ & $0.3297 \pm 0.0700$\\
         \hline
         $h$        & [0.4, 1.0] & $0.7057 \pm 0.1755$ & $0.6430 \pm 0.1427$ & $0.6837 \pm 0.1690$ & $0.7001 \pm 0.1667$ & $0.6967 \pm 0.1732$ & $0.7125 \pm 0.1585$\\
         \hline
         $w_0$ & [-2.0, 0.0] & - & $-1.0960 \pm 0.2087$ & - & $-0.9218 \pm 0.1372$ & - & $-1.0168 \pm 0.0679$\\
         \hline
         $w_a$ &  [-2.0, 2.0] & - & $-0.0996 \pm 1.0031$  & - & $-0.0666 \pm 0.8226$ & - & $-0.4305 \pm 0.8789$\\
         \hline
          $-2\log \Like_{max}$  & - & 1024.9833 & 1024.9165 & 1403.1125 & 1402.6494 & 5502.7526 &  5501.2499\\
         \hline
         $\log Z$ & - & -515.9625 & -517.4652 & -705.2655 & -706.8580 & -2756.3249 & -2759.0397\\
         \hline
         $\log B_{\Lambda CDM, CPL}$ &  & - & 1.5027 & - & 1.5925 & - & 2.7148 \\
         \hline
         Evidence & - & - & \changes{Moderate} & - & \changes{Moderate} & - & \changes{Moderate} \\
         \hline
    \end{tabular}
    \caption{Bayesian analysis to $\Lambda$CDM and CPL models using Pantheon, Pantheon+ and LSST data. We can notice the priors of their parameters, their parameter estimation, $-2\log \Like_{max}$, the log-Bayesian evidence $\log Z$, the log-Bayes factor  $\log B$ of CPL and $\Lambda$CDM for each dataset. Log-Bayes factor is defined as $\log B_{ab}\equiv \log Z_a - \log Z_b$, and according to Jeffrey's scale \cite{vazquez2012bayesian} we can consider weak evidence if $0 \leq |\Delta\log Z|  < 1$, moderate evidence if $1 \leq | \Delta\log Z|  < 3$, strong evidence if $3 \leq | \Delta\log Z|  < 5$, and decisive evidence if $| \Delta\log \mathcal{Z} | \geq 5$, in favour of the preferred model; if $\Delta\log Z$ is negative it is in favour of the second term, and if it is positive the first term is favoured \cite{hee2016bayesian}. In our results, $\Lambda$CDM is always the preferred model.}
    \label{tab:bayesian}
\end{table*}

\subsection{Neural Reconstruction}

\begin{figure*}[t!]
     \centering
      \makebox[11cm][c]{
            \includegraphics[trim=10mm 0mm 25mm 20mm, clip,width=9cm, height=6.5cm]{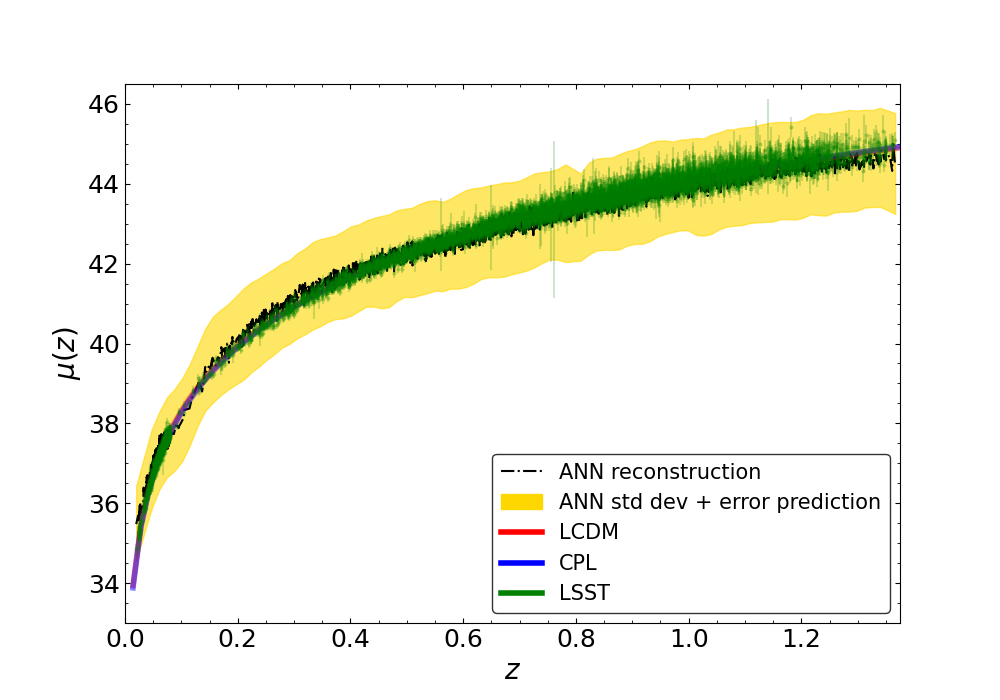}
            \includegraphics[trim=10mm 0mm 25mm 20mm, clip, width=9cm, height=6.5cm]{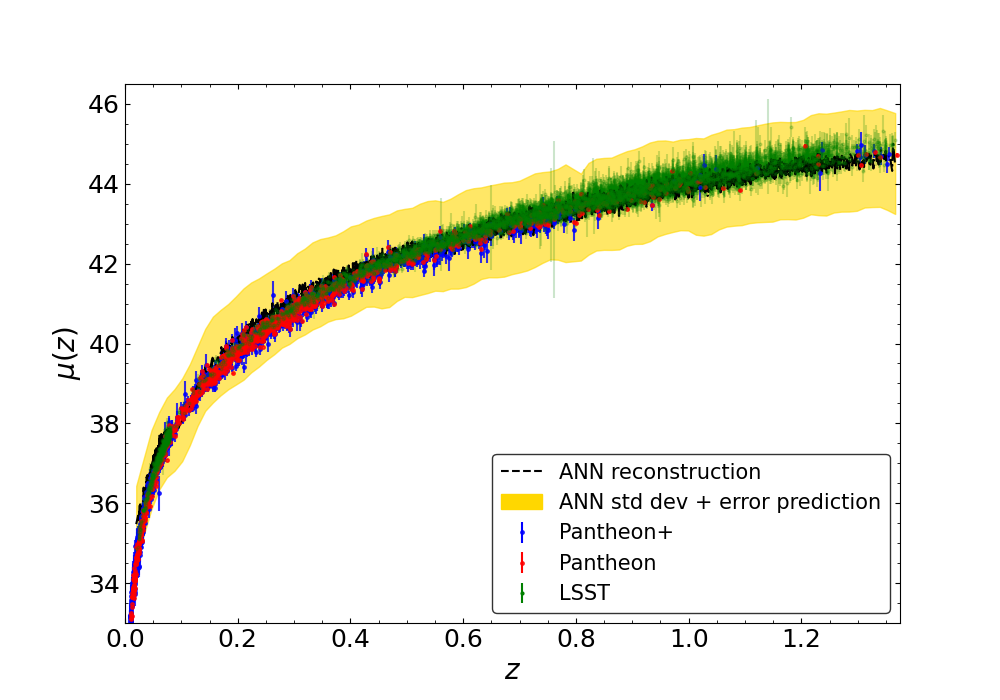}
            }
     \caption{Neural network reconstruction for distance modulus, $\mu(z)$ (black) with the standard deviation of the \changes{Monte Carlo Dropout} predictions plus the modeled error (yellow region) using LSST data (green dots). \textit{Left}: With  $\Lambda$CDM and CPL using the values of the parameter estimation using LSST data. \textit{Right}: In comparison with data points from other SNIa surveys, Pantheon and Pantheon+ using $\Lambda$CDM model only.}
     \label{fig:ann_reconstructions}
 \end{figure*}

Finally, after careful training of a neural network model with the LSST simulated data, we generated the model-independent reconstruction for $\mu(z)$ as can be seen in Fig.~\ref{fig:ann_reconstructions}. In Fig.\ref{fig:ann_reconstructions} we plot the distance modulus ($\mu$) from the datasets of LSST sim and Pantheon, Pantheon+, and then we overlay with the reconstructed distance modulus (black dashed lines) along with their associated uncertainties from the ANN predictions plus the modeled errors (yellow regions). We also plot the theoretically computed $\mu(z)$, \changes{obtained from the Bayesian parameter estimation using the same data,} from the two cosmological dark energy models of $\Lambda$CDM and CPL. From the left panel of Fig.~\ref{fig:ann_reconstructions} it can be seen that the reconstructed data points from the ANN model and the LSST observations (green points) are very close. Next, we computed the MSE for each of these three cases with the observed LSST dataset, shown in Table~\ref{tab:meansq_error}. From Table~\ref{tab:meansq_error} we see that MSE for ANN is slightly better than $\Lambda$CDM or CPL. This implies that the artificial neural network (ANN) based non-parametric reconstruction model can indeed more closely replicate the data than the theoretical models, albeit by a small margin. \changes{Overfitting in the neural reconstruction is avoided by Monte Carlo Dropout, which works as a regularisation method, and with careful tuning of hyperparameters with genetic algorithms. The interesting thing about the neural model is that, although it is a data-driven model without any physical or cosmological assumptions, it is similar to LCDM and CPL.}

The second row in Table~\ref{tab:meansq_error} gives additional insight by comparing the MSE between the ANN reconstructed $\mu(z)$ and the corresponding $\Lambda$CDM or CPL based distances. It is worth noting that, due to the hyperparameter tuning of the ANN and the Monte Carlo dropout, we have a good model based on the data, which looks visually smooth and does not have any overfitting or underfitting in machine learning terms.

\begin{figure*}[h!]
     \centering
     \includegraphics[width=1\textwidth]{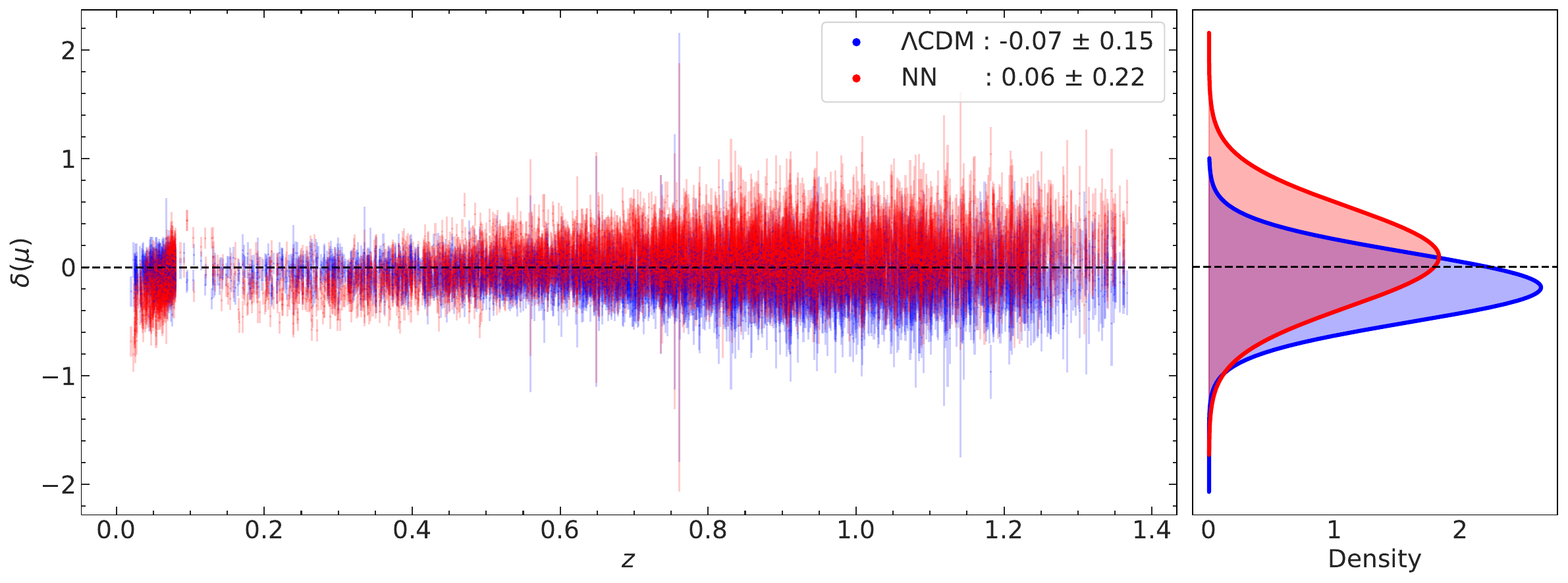}
     \caption{Residual plot ($\delta\mu$) for two cosmological models as shown in figure legend. The residuals are computed between the LSST SNIa sim data (Fig.~\ref{fig:ann_reconstructions}) and model-derived distance estimates. The right panel shows a Gaussian distribution of the corresponding residuals.}
     \label{fig:residual}
 \end{figure*}

In Fig.~\ref{fig:residual} we present residual plots for the distance modulus $\mu(z)$, computed as $\delta\mu\mid_i =\mu_i - \mu_{model}(z_i) $ (where $\mu_i$ corresponds to the $i^{th}$ observation from the LSST sim data) for two cases, a) LSST sim data - $\Lambda$CDM theory and b) LSST sim data - ANN model reconstructed $\mu(z)$ (we compare the $\mu(z)$ residuals between the ANN model and the $\Lambda$CDM model only, as $\Lambda$CDM  is seen to have bigger differences with the ANN reconstruction as seen in Table~\ref{tab:meansq_error}). It can be seen that there is \changes{minor} difference
between the residuals from $\Lambda$CDM and ANN model. 
However, we can notice the sensitivity of the ANN model from outliers, which can make the error margins get bigger.   We performed an additional $F$-test to see if there is any statistically significant difference between the distance modulus residuals from $\Lambda$CDM and ANN model \footnote{The  $F$-test is used to compare the variances of two independent samples to test the hypothesis that they are equal. It is defined as the ratio of the variances, $F = \frac{\sigma_1^2}{\sigma_2^2}$, where $\sigma_1^2$ and $\sigma_2^2$ are the sample variances. This statistic follows the $F$-distribution under the null hypothesis that the variances are equal \cite{snedecor1989statistical}}.  From the $F$-test score, the corresponding $p$ value computed is $p\simeq0$, thereby signifying that the ANN model does not provide any statistically significant improvement over the $\Lambda$CDM estimates.

\begin{table}
\captionsetup{justification=raggedright, singlelinecheck=false, font=footnotesize}
    \centering
    \begin{tabular}{|c|c|c|c|}
         \hline
             & $\Lambda$CDM & CPL  & ANN\\
        \hline
         LSST    & 0.05164 & 0.04708 & 0.03797 \\
         \hline
         Neural predictions & 0.02051 & 0.01881 & -- \\
         \hline 
    \end{tabular}
    \caption{Mean squared error between $\mu(z)$ predictions for (first row) three different models ($\Lambda$CDM, CPL and ANN) with the observed $\mu(z)$ from the LSST SNIa sim dataset and (second row) between the ANN reconstructed, or predicted, distance with the corresponding theoretical estimates from $\Lambda$CDM and CPL\changes{, the differences of the second row are useful to analyze what cosmological model is in more agreement with the data-based neural model.}}.
    \label{tab:meansq_error}
\end{table}

\section{Conclusions}\label{sec:conclude} 
In this paper, we have provided a proof of concept, to \changes{test} a deep learning-based non-parametric dark energy reconstruction model \changes{using SNIa simulated data from LSST}. Which uses the state of the art deep learning tools to find solutions to the open question of finding the most accurate dark energy model. While trying to maximize the benefits one can draw from the huge SNIa datasets that the LSST, is going to provide in a couple of years. In this paper, we show that our reconstructed ANN-based model does perform slightly better than the traditional $\Lambda$CDM and CPL models when tracing the observational data. But it also shows possible areas for future studies. For example, the increased standard deviation observed in the ANN model compared to the theoretical model (Fig.~\ref{fig:residual}) underscores the need for further investigation. This work not only demonstrates the viability of deep learning for cosmological reconstruction analysis but also opens avenues for future research to refine these models for higher precision in understanding dark energy better.

In addition, we notice that the $\Lambda$CDM model is favoured in comparison to CPL using the LSST data, suggesting that there might be room for exploring alternate parameterizations of the dark energy equation of state. In contrast, the MSE of the CPL considering the neural reconstruction is better than $\Lambda$CDM, and considering that despite the lack of interpretability, our neural reconstructions have been rigorously trained to have a generalizable model based only on the data (without underfitting and overfitting), without any cosmological or statistical assumptions, therefore in this CPL shows an advantage over $\Lambda$CDM. These results suggest that it is necessary to incorporate other model comparison techniques to have greater confidence in the preference of the data over the theoretical models, and the nonparametric models may be an opportunity to expand the current repertoire.

We can highlight that the forthcoming LSST data may improve the constraints on parameters of the dark energy models, as we have shown in our Bayesian data analysis. On the other hand, using the parameter estimation for $\Lambda$CDM and CPL with LSST data and comparing it with the neural network reconstruction based on LSST simulations, we have shown that there is another model, smooth and without overfitting, that agrees better with these SNIa data. This indicates that further exploration of new theoretical models together with new statistical and computational models based on the forthcoming cosmological datasets will help to have a broader picture of the directions to follow to understand the behaviour of dark energy, presumably with more machine learning involvement.

\section{Data Availability}

The data used in this work is available from authors upon request. The analysis notebooks are available in \href{https://github.com/igomezv/LSST_DE_neural_reconstruction}{Github}.

\acknowledgments 
AM acknowledges useful discussions with Surhud More and Purba Mukherjee. AM also acknowledges Richard Kessler and the LSST DESC for generating the LSST SNIa simulated dataset. IGV thanks the CONACYT postdoctoral grant.

\medskip

\bibliographystyle{unsrt}
\bibliography{bibliography}

\begin{thebibliography}{100}

\bibitem{riess_1998}
Adam~G. {Riess}, Alexei~V. {Filippenko}, Peter {Challis}, Alejandro
  {Clocchiatti}, Alan {Diercks}, et~al.
\newblock {Observational Evidence from Supernovae for an Accelerating Universe
  and a Cosmological Constant [arXiv:astro-ph/9805201]}.
\newblock {\em Astron. J}, 116(3):1009--1038, September 1998.

\bibitem{Perlmutter:1998np}
S.~Perlmutter et~al.
\newblock {Measurements of $\Omega$ and $\Lambda$ from 42 high redshift
  supernovae}.
\newblock {\em Astrophys. J.}, 517:565--586, 1999.

\bibitem{2008ApJ...686..749K}
M.~{Kowalski} and {others}.
\newblock {Improved Cosmological Constraints from New, Old, and Combined
  Supernova Data Sets}.
\newblock {\em \apj}, 686(2):749--778, October 2008.

\bibitem{Betoule2014}
M.~{Betoule} et~al.
\newblock {Improved cosmological constraints from a joint analysis of the
  SDSS-II and SNLS supernova samples}.
\newblock {\em Astron. Astrophys.}, 568:A22, August 2014.

\bibitem{Planck2016}
{Planck Collaboration} and et. {al.}
\newblock {Planck 2015 results. XIII. Cosmological parameters}.
\newblock {\em Astron. Astrophys.}, 594:A13, September 2016.

\bibitem{2012ApJ...746...85S}
N.~{Suzuki} and et. {al}.
\newblock {The Hubble Space Telescope Cluster Supernova Survey. V. Improving
  the Dark-energy Constraints above z > 1 and Building an Early-type-hosted
  Supernova Sample}.
\newblock {\em Astrophys. J.}, 746(1):85, February 2012.

\bibitem{Carroll:2000fy}
Sean~M. Carroll.
\newblock {The Cosmological constant}.
\newblock {\em Living Rev. Rel.}, 4:1, 2001.

\bibitem{SolaPeracaula:2022hpd}
Joan Sola~Peracaula.
\newblock {The cosmological constant problem and running vacuum in the
  expanding universe}.
\newblock {\em Phil. Trans. Roy. Soc. Lond. A}, 380:20210182, 2022.

\bibitem{Planck2018}
N.~{Aghanim}, Y.~{Akrami}, M.~{Ashdown}, J.~{Aumont}, C.~{Baccigalupi}, et~al.
\newblock {Planck 2018 results. VI. Cosmological parameters}.
\newblock {\em Astron. Astrophys.}, 641:A6, 2020.
\newblock [Erratum: Astron.Astrophys. 652, C4 (2021)].

\bibitem{pantheon_new}
Dillon {Brout}, Dan {Scolnic}, Brodie {Popovic}, Adam~G. {Riess}, Joe {Zuntz},
  Rick {Kessler}, Anthony {Carr}, Tamara~M. {Davis}, Samuel {Hinton}, David
  {Jones}, W.~D'Arcy {Kenworthy}, Erik~R. {Peterson}, Khaled {Said}, Georgie
  {Taylor}, Noor {Ali}, Patrick {Armstrong}, Pranav {Charvu}, Arianna {Dwomoh},
  Antonella {Palmese}, Helen {Qu}, Benjamin~M. {Rose}, Christopher~W. {Stubbs},
  Maria {Vincenzi}, Charlotte~M. {Wood}, Peter~J. {Brown}, Rebecca {Chen}, Ken
  {Chambers}, David~A. {Coulter}, Mi~{Dai}, Georgios {Dimitriadis}, Alexei~V.
  {Filippenko}, Ryan~J. {Foley}, Saurabh~W. {Jha}, Lisa {Kelsey}, Robert~P.
  {Kirshner}, Anais {M{\"o}ller}, Jessie {Muir}, Seshadri {Nadathur}, Yen-Chen
  {Pan}, Armin {Rest}, Cesar {Rojas-Bravo}, Masao {Sako}, Matthew~R. {Siebert},
  Mat {Smith}, Benjamin~E. {Stahl}, and Phil {Wiseman}.
\newblock {The Pantheon+ Analysis: Cosmological Constraints}.
\newblock {\em arXiv e-prints}, page arXiv:2202.04077, February 2022.

\bibitem{Pantheon}
D.~M. Scolnic et~al.
\newblock \emph{The Complete Light-curve Sample of Spectroscopically Confirmed
  SNe Ia from Pan-STARRS1 and Cosmological Constraints from the Combined
  Pantheon Sample}.
\newblock {\em Astrophys. J.}, 859(2):101, 2018.

\bibitem{Riess_2022}
Adam~G. Riess et~al.
\newblock \emph{A Comprehensive Measurement of the Local Value of the Hubble
  Constant with 1 km s$^{-1}$ Mpc$^{-1}$ Uncertainty from the Hubble Space
  Telescope and the SH0ES Team}.
\newblock {\em Astrophys. J. Lett.}, 934(1):L7, 2022.

\bibitem{DiValentino:2021izs}
Eleonora Di~Valentino, Olga Mena, Supriya Pan, Luca Visinelli, Weiqiang Yang,
  Alessandro Melchiorri, David~F. Mota, Adam~G. Riess, and Joseph Silk.
\newblock {In the realm of the Hubble tension\textemdash{}a review of
  solutions}.
\newblock {\em Class. Quant. Grav.}, 38(15):153001, 2021.

\bibitem{Aluri:2022hzs}
Pavan~Kumar Aluri et~al.
\newblock {Is the observable Universe consistent with the cosmological
  principle?}
\newblock {\em Class. Quant. Grav.}, 40(9):094001, 2023.

\bibitem{Perivolaropoulos:2021jda}
Leandros Perivolaropoulos and Foteini Skara.
\newblock {Challenges for \ensuremath{\Lambda}CDM: An update}.
\newblock {\em New Astron. Rev.}, 95:101659, 2022.

\bibitem{Kamionkowski:2022pkx}
Marc Kamionkowski and Adam~G. Riess.
\newblock {The Hubble Tension and Early Dark Energy}.
\newblock 11 2022.

\bibitem{Krishnan:2021dyb}
Chethan Krishnan, Roya Mohayaee, Eoin~\'O. Colg\'ain, M.~M. Sheikh-Jabbari, and
  Lu~Yin.
\newblock {Does Hubble tension signal a breakdown in FLRW cosmology?}
\newblock {\em Class. Quant. Grav.}, 38(18):184001, 2021.

\bibitem{Abdalla:2022yfr}
Elcio Abdalla et~al.
\newblock {Cosmology intertwined: A review of the particle physics,
  astrophysics, and cosmology associated with the cosmological tensions and
  anomalies}.
\newblock {\em JHEAp}, 34:49--211, 2022.

\bibitem{Krishnan:2021jmh}
Chethan Krishnan, Roya Mohayaee, Eoin~\'O. Colg\'ain, M.~M. Sheikh-Jabbari, and
  L.~Yin.
\newblock {Hints of FLRW breakdown from supernovae}.
\newblock {\em Phys. Rev. D}, 105(6):063514, 2022.

\bibitem{Smith:2022hwi}
Tristan~L. Smith, Matteo Lucca, Vivian Poulin, Guillermo~F. Abellan, Lennart
  Balkenhol, Karim Benabed, Silvia Galli, and Riccardo Murgia.
\newblock {Hints of early dark energy in Planck, SPT, and ACT data: New physics
  or systematics?}
\newblock {\em Phys. Rev. D}, 106(4):043526, 2022.

\bibitem{sahni2002cosmological}
Varun Sahni.
\newblock The cosmological constant problem and quintessence.
\newblock {\em Class. Quantum Gravity}, 19(13):3435, 2002.

\bibitem{wasserman2006all}
Larry Wasserman.
\newblock {\em All of nonparametric statistics}.
\newblock Springer Science \& Business Media, 2006.

\bibitem{sharma2020reconstruction}
Ranbir Sharma, Ankan Mukherjee, and HK~Jassal.
\newblock Reconstruction of late-time cosmology using principal component
  analysis.
\newblock {\em preprint (\href{arXiv:2004.01393}{arXiv:2004.01393})}, 2020.

\bibitem{r8}
Francesca Gerardi, Matteo Martinelli, and Alessandra Silvestri.
\newblock {Reconstruction of the Dark Energy equation of state from latest
  data: the impact of theoretical priors}.
\newblock {\em JCAP}, 07:042, 2019.

\bibitem{williams2006gaussian}
Christopher~KI Williams and Carl~Edward Rasmussen.
\newblock {\em Gaussian processes for machine learning}, volume~2.
\newblock MIT press Cambridge, MA, 2006.

\bibitem{Keeley:2020aym}
Ryan~E. Keeley, Arman Shafieloo, Gong-Bo Zhao, Jose~Alberto Vazquez, and
  Hanwool Koo.
\newblock Reconstructing the universe: Testing the mutual consistency of the
  pantheon and {SDSS}/{eBOSS} {BAO} data sets with gaussian processes.
\newblock {\em The Astronomical Journal}, 161(3):151, Feb 2021.

\bibitem{l2020defying}
Benjamin L’Huillier, Arman Shafieloo, David Polarski, and Alexei~A
  Starobinsky.
\newblock Defying the laws of gravity i: model-independent reconstruction of
  the universe expansion from growth data.
\newblock {\em Monthly Notices of the Royal Astronomical Society},
  494(1):819--826, 2020.

\bibitem{r10}
Purba Mukherjee and Narayan Banerjee.
\newblock {Revisiting a non-parametric reconstruction of the deceleration
  parameter from combined background and the growth rate data}.
\newblock {\em Phys. Dark Univ.}, 36:100998, 2022.

\bibitem{escamilla2023model}
Luis~A Escamilla, {\"O}zg{\"u}r Akarsu, Eleonora Di~Valentino, and J~Alberto
  Vazquez.
\newblock Model-independent reconstruction of the interacting dark energy
  kernel: Binned and gaussian process.
\newblock {\em Journal of Cosmology and Astroparticle Physics}, 2023(11):051,
  2023.

\bibitem{montiel2014nonparametric}
Ariadna Montiel, Ruth Lazkoz, Irene Sendra, Celia Escamilla-Rivera, and
  Vincenzo Salzano.
\newblock Nonparametric reconstruction of the cosmic expansion with local
  regression smoothing and simulation extrapolation.
\newblock {\em Physical Review D}, 89(4):043007, 2014.

\bibitem{sahni2006reconstructing}
Varun Sahni and Alexei Starobinsky.
\newblock Reconstructing dark energy.
\newblock {\em Int. J. Mod. Phys. D}, 15(12):2105--2132, 2006.

\bibitem{holsclaw2010nonparametric}
Tracy Holsclaw, Ujjaini Alam, Bruno Sanso, Herbert Lee, Katrin Heitmann, Salman
  Habib, and David Higdon.
\newblock Nonparametric dark energy reconstruction from supernova data.
\newblock {\em Physical Review Letters}, 105(24):241302, 2010.

\bibitem{zhao2017dynamical}
Gong-Bo Zhao, Marco Raveri, Levon Pogosian, Yuting Wang, Robert~G Crittenden,
  Will~J Handley, Will~J Percival, Florian Beutler, Jonathan Brinkmann,
  Chia-Hsun Chuang, et~al.
\newblock Dynamical dark energy in light of the latest observations.
\newblock {\em Nature Astronomy}, 1(9):627--632, 2017.

\bibitem{gomez2023neuralrecs}
Isidro G{\'o}mez-Vargas, Ricardo Medel-Esquivel, Ricardo Garc{\'\i}a-Salcedo,
  and J~Alberto V{\'a}zquez.
\newblock Neural network reconstructions for the hubble parameter, growth rate
  and distance modulus.
\newblock {\em The European Physical Journal C}, 83(4):304, 2023.

\bibitem{wei2017improved}
Jun-Jie Wei and Xue-Feng Wu.
\newblock An improved method to measure the cosmic curvature.
\newblock {\em The Astrophysical Journal}, 838(2):160, 2017.

\bibitem{lin2019non}
Hai-Nan Lin, Xin Li, and Li~Tang.
\newblock Non-parametric reconstruction of dark energy and cosmic expansion
  from the pantheon compilation of type ia supernovae.
\newblock {\em Chinese Phys. C}, 43(7):075101, 2019.

\bibitem{wang2020reconstructing}
Guo-Jian Wang, Xiao-Jiao Ma, Si-Yao Li, and Jun-Qing Xia.
\newblock Reconstructing functions and estimating parameters with artificial
  neural networks: A test with a hubble parameter and sne ia.
\newblock {\em Astrophysical Journal Supplement Series}, 246(1):13, 2020.

\bibitem{escamilla2020deep}
Celia Escamilla-Rivera, Maryi A~Carvajal Quintero, and Salvatore Capozziello.
\newblock A deep learning approach to cosmological dark energy models.
\newblock {\em Journal of Cosmology and Astroparticle Physics}, 2020(03):008,
  2020.

\bibitem{dialektopoulos2023neural}
Konstantinos~F Dialektopoulos, Purba Mukherjee, Jackson~Levi Said, and Jurgen
  Mifsud.
\newblock Neural network reconstruction of cosmology using the pantheon
  compilation.
\newblock {\em The European Physical Journal C}, 83(10):1--13, 2023.

\bibitem{arjona2020machine}
Rub{\'e}n Arjona.
\newblock Machine learning meets the redshift evolution of the cmb temperature.
\newblock {\em Journal of Cosmology and Astroparticle Physics}, 2020(08):009,
  2020.

\bibitem{arjona2020can}
Rub{\'e}n Arjona and Savvas Nesseris.
\newblock What can machine learning tell us about the background expansion of
  the universe?
\newblock {\em Physical Review D}, 101(12):123525, 2020.

\bibitem{arjona2020hints}
Rub{\'e}n Arjona and Savvas Nesseris.
\newblock Hints of dark energy anisotropic stress using machine learning.
\newblock {\em Journal of Cosmology and Astroparticle Physics}, 2020(11):042,
  2020.

\bibitem{arjona2022testing}
Rub{\'e}n Arjona, Alessandro Melchiorri, and Savvas Nesseris.
\newblock Testing the $\lambda$cdm paradigm with growth rate data and machine
  learning.
\newblock {\em Journal of Cosmology and Astroparticle Physics}, 2022(05):047.
  [arXiv:2107.04343], 2022.

\bibitem{lin2017does}
Henry~W Lin, Max Tegmark, and David Rolnick.
\newblock Why does deep and cheap learning work so well?
\newblock {\em J. Statistical Physics}, 168(6):1223--1247, 2017.

\bibitem{peel2019distinguishing}
Austin Peel, Florian Lalande, Jean-Luc Starck, Valeria Pettorino, Julian
  Merten, Carlo Giocoli, Massimo Meneghetti, and Marco Baldi.
\newblock Distinguishing standard and modified gravity cosmologies with machine
  learning.
\newblock {\em Physical Review D}, 100(2):023508, 2019.

\bibitem{wang2020machine}
Guo-Jian Wang, Xiao-Jiao Ma, and Jun-Qing Xia.
\newblock Machine learning the cosmic curvature in a model-independent way.
\newblock {\em Monthly Notices of the Royal Astronomical Society},
  501(4):5714--5722, 2021.

\bibitem{gomez2023neuralgenetic}
Isidro G{\'o}mez-Vargas, Joshua~Briones Andrade, and J~Alberto V{\'a}zquez.
\newblock Neural networks optimized by genetic algorithms in cosmology.
\newblock {\em Physical Review D}, 107(4):043509, 2023.

\bibitem{chacon2023analysis}
Jazhiel Chac{\'o}n, Isidro G{\'o}mez-Vargas, Ricardo~Menchaca M{\'e}ndez, and
  J~Alberto V{\'a}zquez.
\newblock Analysis of dark matter halo structure formation in n-body
  simulations with machine learning.
\newblock {\em Physical Review D}, 107(12):123515, 2023.

\bibitem{medel2023cosmological}
Ricardo Medel-Esquivel, Isidro G{\'o}mez-Vargas, Alejandro~A
  Morales~S{\'a}nchez, Ricardo Garc{\'\i}a-Salcedo, and Jos{\'e}
  Alberto~V{\'a}zquez.
\newblock Cosmological parameter estimation with genetic algorithms.
\newblock {\em Universe}, 10(1):11, 2023.

\bibitem{dieleman2015rotation}
Sander Dieleman, Kyle~W Willett, and Joni Dambre.
\newblock Rotation-invariant convolutional neural networks for galaxy
  morphology prediction.
\newblock {\em Monthly Notices of the Royal Astronomical Society},
  450(2):1441--1459, 2015.

\bibitem{ntampaka2019deep}
Michelle Ntampaka, J~ZuHone, D~Eisenstein, D~Nagai, A~Vikhlinin, L~Hernquist,
  F~Marinacci, D~Nelson, R~Pakmor, A~Pillepich, et~al.
\newblock A deep learning approach to galaxy cluster x-ray masses.
\newblock {\em The Astrophysical Journal}, 876(1):82, 2019.

\bibitem{rodriguez2018fast}
Andres~C Rodr{\'\i}guez, Tomasz Kacprzak, Aurelien Lucchi, Adam Amara, Raphael
  Sgier, Janis Fluri, Thomas Hofmann, and Alexandre R{\'e}fr{\'e}gier.
\newblock Fast cosmic web simulations with generative adversarial networks.
\newblock {\em Comp. Astrophys. and Cosmology}, 5(1):4, 2018.

\bibitem{he2019learning}
Siyu He, Yin Li, Yu~Feng, Shirley Ho, Siamak Ravanbakhsh, Wei Chen, and
  Barnab{\'a}s P{\'o}czos.
\newblock Learning to predict the cosmological structure formation.
\newblock {\em Proc. Natl. Acad. Sci.}, 116(28):13825--13832, 2019.

\bibitem{alsing2019fast}
Justin Alsing, Tom Charnock, Stephen Feeney, and Benjamin Wandelt.
\newblock Fast likelihood-free cosmology with neural density estimators and
  active learning.
\newblock {\em Monthly Notices of the Royal Astronomical Society},
  488(3):4440--4458, 2019.

\bibitem{li2019model}
Shi-Yu Li, Yun-Long Li, and Tong-Jie Zhang.
\newblock Model comparison of dark energy models using deep network.
\newblock {\em Res. Astron. Astrophys.}, 19(9):137, 2019.

\bibitem{hortua2020constraining}
H{\'e}ctor~J Hort{\'u}a, Luigi Malag{\`o}, and Riccardo Volpi.
\newblock Constraining the reionization history using bayesian normalizing
  flows.
\newblock {\em Machine Learning: Science and Technology}, 1(3):035014, 2020.

\bibitem{hortua2020parameter}
H{\'e}ctor~J Hort{\'u}a, Riccardo Volpi, Dimitri Marinelli, and Luigi
  Malag{\`o}.
\newblock Parameter estimation for the cosmic microwave background with
  bayesian neural networks.
\newblock {\em Physical Review D}, 102(10):103509, 2020.

\bibitem{FoM}
Andreas Albrecht et~al.
\newblock {Report of the Dark Energy Task Force}.
\newblock {\em arXiv:astro-ph/0609591}, 9 2006.

\bibitem{EuclidTheoryWorkingGroup:2012gxx}
Luca Amendola et~al.
\newblock {Cosmology and fundamental physics with the Euclid satellite}.
\newblock {\em Living Rev. Rel.}, 16:6, 2013.

\bibitem{https://doi.org/10.48550/arxiv.0912.0914}
A.~Cimatti, R.~Laureijs, B.~Leibundgut, S.~Lilly, R.~Nichol, A.~Refregier,
  P.~Rosati, M.~Steinmetz, N.~Thatte, and E.~Valentijn.
\newblock {Euclid Assessment Study Report for the ESA Cosmic Visions}.
\newblock {\em arXiv:0912.0914}, 12 2009.

\bibitem{LSSTDarkEnergyScience:2018jkl}
Rachel Mandelbaum et~al.
\newblock {The LSST Dark Energy Science Collaboration (DESC) Science
  Requirements Document}.
\newblock {\em arXiv:1809.01669}, 9 2018.

\bibitem{Zhan:2017uwu}
Hu~Zhan and J.~Anthony Tyson.
\newblock {Cosmology with the Large Synoptic Survey Telescope: an Overview}.
\newblock {\em Rept. Prog. Phys.}, 81(6):066901, 2018.

\bibitem{LSST:2008ijt}
\v{Z}eljko Ivezi\'c et~al.
\newblock {LSST: from Science Drivers to Reference Design and Anticipated Data
  Products}.
\newblock {\em Astrophys. J.}, 873(2):111, 2019.

\bibitem{Akeson:2019biv}
Rachel Akeson et~al.
\newblock {The Wide Field Infrared Survey Telescope: 100 Hubbles for the
  2020s}.
\newblock {\em arXiv:1902.05569}, 2 2019.

\bibitem{TMT}
Warren Skidmore et~al.
\newblock {Thirty Meter Telescope Detailed Science Case: 2015}.
\newblock {\em Res. Astron. Astrophys.}, 15(12):1945--2140, 2015.

\bibitem{Gardner_2006}
Jonathan~P. Gardner, John~C. Mather, Mark Clampin, Rene Doyon, Matthew~A.
  Greenhouse, Heidi~B. Hammel, John~B. Hutchings, Peter Jakobsen, Simon~J.
  Lilly, Knox~S. Long, Jonathan~I. Lunine, Mark~J. Mccaughrean, Matt Mountain,
  John Nella, George~H. Rieke, Marcia~J. Rieke, Hans-Walter Rix, Eric~P. Smith,
  George Sonneborn, Massimo Stiavelli, H.~S. Stockman, Rogier~A. Windhorst, and
  Gillian~S. Wright.
\newblock The james webb space telescope.
\newblock {\em Space Science Reviews}, 123(4):485--606, 4 2006.

\bibitem{plasticc_announce}
Jr. {Allam}, Tarek, Anita {Bahmanyar}, Rahul {Biswas}, Mi~{Dai}, Llu{\'\i}s
  {Galbany}, Ren{\'e}e {Hlo{\v{z}}ek}, Emille E.~O. {Ishida}, Saurabh~W. {Jha},
  David~O. {Jones}, Richard {Kessler}, Michelle {Lochner}, Ashish~A. {Mahabal},
  Alex~I. {Malz}, Kaisey~S. {Mandel}, Juan~Rafael {Mart{\'\i}nez-Galarza},
  Jason~D. {McEwen}, Daniel {Muthukrishna}, Gautham {Narayan}, Hiranya
  {Peiris}, Christina~M. {Peters}, Kara {Ponder}, Christian~N. {Setzer}, {The
  LSST Dark Energy Science Collaboration}, The {LSST Transients}, and {Variable
  Stars Science Collaboration}.
\newblock {The Photometric LSST Astronomical Time-series Classification
  Challenge (PLAsTiCC): Data set}.
\newblock {\em arXiv:1810.00001}, September 2018.

\bibitem{eos1}
Eric~V. Linder.
\newblock {Exploring the expansion history of the universe [arXiv:0208512]}.
\newblock {\em Phys. Rev. Lett.}, 90:091301, 2003.

\bibitem{eos2}
Bharat {Ratra} and P.~J.~E. {Peebles}.
\newblock {Cosmological consequences of a rolling homogeneous scalar field}.
\newblock {\em Phys. Rev. D}, 37(12):3406--3427, June 1988.

\bibitem{2017ApJ...850..183Z}
Zhongxu {Zhai}, Michael {Blanton}, An{\v{z}}e {Slosar}, and Jeremy {Tinker}.
\newblock {An Evaluation of Cosmological Models from the Expansion and Growth
  of Structure Measurements}.
\newblock {\em \apj}, 850(2):183, December 2017.

\bibitem{1960ApJ...132..565H}
F.~{Hoyle} and William~A. {Fowler}.
\newblock {Nucleosynthesis in Supernovae.}
\newblock {\em \apj}, 132:565, November 1960.

\bibitem{Trip1998}
Robert {Tripp}.
\newblock {A two-parameter luminosity correction for Type IA supernovae}.
\newblock {\em Astron. Astrophys.}, 331:815--820, March 1998.

\bibitem{bbc}
R.~{Kessler} and D.~{Scolnic}.
\newblock {Correcting Type Ia Supernova Distances for Selection Biases and
  Contamination in Photometrically Identified Samples}.
\newblock {\em Astrophys. J.}, 836(1):56, February 2017.

\bibitem{panstarrs}
A.~Rest et~al.
\newblock {Cosmological Constraints from Measurements of Type Ia Supernovae
  discovered during the first 1.5 yr of the Pan-STARRS1 Survey}.
\newblock {\em Astrophys. J.}, 795(1):44, 2014.

\bibitem{s1}
M.~{Sullivan} et~al.
\newblock {SNLS3: Constraints on Dark Energy Combining the Supernova Legacy
  Survey Three-year Data with Other Probes [arXiv:1104.1444]}.
\newblock {\em Astrophys. J.}, 737(2):102, August 2011.

\bibitem{Dhawan:2021ztf}
S.~Dhawan et~al.
\newblock {The Zwicky Transient Facility Type Ia supernova survey: first data
  release and results}.
\newblock {\em Mon. Not. Roy. Astron. Soc.}, 510(2):2228--2241, 2022.

\bibitem{Kessler2010}
Richard {Kessler}, David {Cinabro}, Bruce {Bassett}, Benjamin {Dilday},
  Joshua~A. {Frieman}, et~al.
\newblock {Photometric Estimates of Redshifts and Distance Moduli for Type Ia
  Supernovae}.
\newblock {\em Astrophys. J.}, 717(1):40--57, July 2010.

\bibitem{Linder2019}
Eric~V. {Linder} and Ayan {Mitra}.
\newblock {Photometric supernovae redshift systematics requirements
  [arXiv:1907.00985]}.
\newblock {\em Phys. Rev. D}, 100(4):043542, August 2019.

\bibitem{Mitra2021}
Ayan Mitra and Eric~V. Linder.
\newblock {Cosmology requirements on supernova photometric redshift systematics
  for the Rubin LSST and Roman Space Telescope}.
\newblock {\em Phys. Rev. D}, 103(2):023524, 2021.

\bibitem{re1}
Shin'ichi Nojiri and Sergei~D. Odintsov.
\newblock {Introduction to modified gravity and gravitational alternative for
  dark energy}.
\newblock {\em eConf}, C0602061:06, 2006.

\bibitem{re2}
Martin Sahlen, Andrew~R Liddle, and David Parkinson.
\newblock {Quintessence reconstructed: New constraints and tracker viability}.
\newblock {\em Phys. Rev. D}, 75:023502, 2007.

\bibitem{re3}
Chao Li, Daniel~E. Holz, and Asantha Cooray.
\newblock {Direct Reconstruction of the Dark Energy Scalar-Field Potential}.
\newblock {\em Phys. Rev. D}, 75:103503, 2007.

\bibitem{re4}
Shin'ichi Nojiri and Sergei~D. Odintsov.
\newblock {Modified gravity and its reconstruction from the universe expansion
  history}.
\newblock {\em J. Phys. Conf. Ser.}, 66:012005, 2007.

\bibitem{re5}
Caroline Zunckel and R.~Trotta.
\newblock {Reconstructing the history of dark energy using maximum entropy}.
\newblock {\em Mon. Not. Roy. Astron. Soc.}, 380:865, 2007.

\bibitem{re6}
Arman Shafieloo.
\newblock {Model Independent Reconstruction of the Expansion History of the
  Universe and the Properties of Dark Energy}.
\newblock {\em Mon. Not. Roy. Astron. Soc.}, 380:1573--1580, 2007.

\bibitem{re7}
Eric~V. Linder.
\newblock {The Dynamics of Quintessence, The Quintessence of Dynamics}.
\newblock {\em Gen. Rel. Grav.}, 40:329--356, 2008.

\bibitem{Liddle2007}
A.~R. Liddle.
\newblock Information criteria for astrophysical model selection.
\newblock {\em Monthly Notices of the Royal Astronomical Society: Letters},
  377(1):L74--L78, 2007.

\bibitem{Copeland2006}
E.~J. Copeland, M.~Sami, and S.~Tsujikawa.
\newblock Dynamics of dark energy.
\newblock {\em International Journal of Modern Physics D}, 15(11):1753--1936,
  2006.

\bibitem{Trotta2008}
R.~Trotta.
\newblock Bayes in the sky: Bayesian inference and model selection in
  cosmology.
\newblock {\em Contemporary Physics}, 49(2):71--104, 2008.

\bibitem{r9}
Benjamin L'Huillier, Arman Shafieloo, David Polarski, and Alexei~A.
  Starobinsky.
\newblock {Defying the laws of Gravity I: Model-independent reconstruction of
  the Universe expansion from growth data}.
\newblock {\em Mon. Not. Roy. Astron. Soc.}, 494(1):819--826, 2020.

\bibitem{r11}
Alexander Bonilla, Suresh Kumar, and Rafael~C. Nunes.
\newblock {Measurements of $H_0$ and reconstruction of the dark energy
  properties from a model-independent joint analysis}.
\newblock {\em Eur. Phys. J. C}, 81(2):127, 2021.

\bibitem{r12}
Luis~A. Escamilla and J.~Alberto Vazquez.
\newblock {Model selection applied to reconstructions of the Dark Energy}.
\newblock {\em Eur. Phys. J. C}, 83(3):251, 2023.

\bibitem{r13}
Purba Mukherjee.
\newblock {\em {Non-parametric Reconstruction Of Some Cosmological
  Parameters}}.
\newblock PhD thesis, IISER, Kolkata, 2022.

\bibitem{r14}
Marina Seikel, Chris Clarkson, and Mathew Smith.
\newblock {Reconstruction of dark energy and expansion dynamics using Gaussian
  processes}.
\newblock {\em JCAP}, 06:036, 2012.

\bibitem{Wang:2020sxl}
Yu-Chen Wang, Yuan-Bo Xie, Tong-Jie Zhang, Hui-Chao Huang, Tingting Zhang, and
  Kun Liu.
\newblock {Likelihood-free Cosmological Constraints with Artificial Neural
  Networks: An Application on Hubble Parameters and SNe Ia}.
\newblock {\em Astrophys. J. Supp.}, 254(2):43, 2021.

\bibitem{Wang:2019vxv}
Guo-Jian Wang, Xiao-Jiao Ma, Si-Yao Li, and Jun-Qing Xia.
\newblock {Reconstructing Functions and Estimating Parameters with Artificial
  Neural Networks: A Test with a Hubble Parameter and SNe Ia}.
\newblock {\em Astrophys. J. Suppl.}, 246(1):13, 2020.

\bibitem{hornik1990universal}
Kurt Hornik, Maxwell Stinchcombe, and Halbert White.
\newblock Universal approximation of an unknown mapping and its derivatives
  using multilayer feedforward networks.
\newblock {\em Neural networks}, 3(5):551--560, 1990.

\bibitem{dialektopoulos2022neural}
Konstantinos Dialektopoulos, Jackson~Levi Said, Jurgen Mifsud, Joseph Sultana,
  and Kristian~Zarb Adami.
\newblock Neural network reconstruction of late-time cosmology and null tests.
\newblock {\em Journal of Cosmology and Astroparticle Physics}, 2022(02):023.
  [arXiv:2111.11462], 2022.

\bibitem{goodfellow2016deep}
Ian Goodfellow, Yoshua Bengio, Aaron Courville, and Yoshua Bengio.
\newblock {\em Deep learning}, volume~1.
\newblock MIT press Cambridge, 2016.

\bibitem{bishop2006pattern}
Christopher~M. Bishop.
\newblock {\em Pattern Recognition and Machine Learning (Information Science
  and Statistics)}.
\newblock Springer-Verlag, Berlin, Heidelberg, 2006.

\bibitem{nielsen2015neural}
Michael~A. Nielsen.
\newblock Neural networks and deep learning, 2018.

\bibitem{Cahn2009}
Robert~N. {Cahn}.
\newblock {\em {Dark Energy Task Force}}, pages 685--695.
\newblock World Scientific, 2009.

\bibitem{ivezic}
\v{Z}eljko Ivezi\'c et~al.
\newblock {LSST: from Science Drivers to Reference Design and Anticipated Data
  Products}.
\newblock {\em Astrophys. J.}, 873(2):111, 2019.

\bibitem{dc2}
{LSST Dark Energy Science Collaboration (LSST DESC)}, Bela {Abolfathi}, David
  {Alonso}, Robert {Armstrong}, {\'E}ric {Aubourg}, Humna {Awan}, Yadu~N.
  {Babuji}, Franz~Erik {Bauer}, Rachel {Bean}, George {Beckett}, Rahul
  {Biswas}, Joanne~R. {Bogart}, Dominique {Boutigny}, Kyle {Chard}, James
  {Chiang}, Chuck~F. {Claver}, Johann {Cohen-Tanugi}, C{\'e}line {Combet},
  Andrew~J. {Connolly}, Scott~F. {Daniel}, Seth~W. {Digel}, Alex
  {Drlica-Wagner}, Richard {Dubois}, Emmanuel {Gangler}, Eric {Gawiser}, Thomas
  {Glanzman}, Phillipe {Gris}, Salman {Habib}, Andrew~P. {Hearin}, Katrin
  {Heitmann}, Fabio {Hernandez}, Ren{\'e}e {Hlo{\v{z}}ek}, Joseph {Hollowed},
  Mustapha {Ishak}, {\v{Z}}eljko {Ivezi{\'c}}, Mike {Jarvis}, Saurabh~W. {Jha},
  Steven~M. {Kahn}, J.~Bryce {Kalmbach}, Heather~M. {Kelly}, Eve {Kovacs},
  Danila {Korytov}, K.~Simon {Krughoff}, Craig~S. {Lage}, Fran{\c{c}}ois
  {Lanusse}, Patricia {Larsen}, Laurent {Le Guillou}, Nan {Li}, Emily~Phillips
  {Longley}, Robert~H. {Lupton}, Rachel {Mandelbaum}, Yao-Yuan {Mao}, Phil
  {Marshall}, Joshua~E. {Meyers}, Marc {Moniez}, Christopher~B. {Morrison},
  Andrei {Nomerotski}, Paul {O'Connor}, HyeYun {Park}, Ji~Won {Park}, Julien
  {Peloton}, Daniel {Perrefort}, James {Perry}, St{\'e}phane {Plaszczynski},
  Adrian {Pope}, Andrew {Rasmussen}, Kevin {Reil}, Aaron~J. {Roodman}, Eli~S.
  {Rykoff}, F.~Javier {S{\'a}nchez}, Samuel~J. {Schmidt}, Daniel {Scolnic},
  Christopher~W. {Stubbs}, J.~Anthony {Tyson}, Thomas~D. {Uram}, Antonio
  {Villarreal}, Christopher~W. {Walter}, Matthew~P. {Wiesner}, W.~Michael
  {Wood-Vasey}, and Joe {Zuntz}.
\newblock {The LSST DESC DC2 Simulated Sky Survey}.
\newblock {\em Astrophys. J., Suppl. Ser.}, 253(1):31, March 2021.

\bibitem{Sanchez}
B.~{S{\'a}nchez}, R.~{Kessler}, D.~{Scolnic}, B.~{Armstrong}, R.~{Biswas},
  J.~{Bogart}, J.~{Chiang}, J.~{Cohen-Tanugi}, D.~{Fouchez}, Ph. {Gris},
  K.~{Heitmann}, R.~{Hlo{\v{z}}ek}, S.~{Jha}, H.~{Kelly}, S.~{Liu},
  G.~{Narayan}, B.~{Racine}, E.~{Rykoff}, M.~{Sullivan}, C.~{Walter},
  M.~{Wood-Vasey}, and {The LSST Dark Energy Science Collaboration}.
\newblock {SNIa-Cosmology Analysis Results from Simulated LSST Images: from
  Difference Imaging to Constraints on Dark Energy}.
\newblock {\em arXiv e-prints}, page arXiv:2111.06858, November 2021.

\bibitem{mitra2023using}
Ayan Mitra, Richard Kessler, Surhud More, Renee Hlozek, LSST Dark
  Energy~Science Collaboration, et~al.
\newblock Using host galaxy photometric redshifts to improve cosmological
  constraints with type ia supernovae in the lsst era.
\newblock {\em The Astrophysical Journal}, 944(2):212, 2023.

\bibitem{snana}
Richard {Kessler}, Joseph~P. {Bernstein}, David {Cinabro}, Benjamin {Dilday},
  Joshua~A. {Frieman}, Saurabh {Jha}, Stephen {Kuhlmann}, Gajus {Miknaitis},
  Masao {Sako}, Matt {Taylor}, and Jake {Vanderplas}.
\newblock {SNANA: A Public Software Package for Supernova Analysis}.
\newblock {\em Publ. Astron. Soc. Pac.}, 121(883):1028, sep 2009.

\bibitem{salt}
Julien Guy et~al.
\newblock {SALT: A Spectral adaptive Light curve Template for Type Ia
  supernovae}.
\newblock {\em Astron. Astrophys.}, 443:781--791, 2005.

\bibitem{Graham2018_photoz}
Melissa~L. {Graham} et~al.
\newblock {Photometric Redshifts with the LSST. II. The Impact of Near-infrared
  and Near-ultraviolet Photometry [arXiv:2004.07885]}.
\newblock {\em Astron. J}, 159(6):258, June 2020.

\bibitem{4MOST2}
R.~S. {de Jong}, O.~{Agertz}, A.~A. {Berbel}, J.~{Aird}, D.~A. {Alexander},
  et~al.
\newblock {4MOST: Project overview and information for the First Call for
  Proposals}.
\newblock {\em The Messenger}, 175:3--11, March 2019.

\bibitem{scolnic2018complete}
Daniel~Moshe Scolnic, DO~Jones, A~Rest, YC~Pan, R~Chornock, RJ~Foley, ME~Huber,
  R~Kessler, Gautham Narayan, AG~Riess, et~al.
\newblock The complete light-curve sample of spectroscopically confirmed sne ia
  from pan-starrs1 and cosmological constraints from the combined pantheon
  sample.
\newblock {\em The Astrophysical Journal}, 859(2):101, 2018.

\bibitem{brout2022pantheon+}
Dillon Brout, Dan Scolnic, Brodie Popovic, Adam~G Riess, Anthony Carr, Joe
  Zuntz, Rick Kessler, Tamara~M Davis, Samuel Hinton, David Jones, et~al.
\newblock The pantheon+ analysis: cosmological constraints.
\newblock {\em The Astrophysical Journal}, 938(2):110, 2022.

\bibitem{2002SPIE.4836..154K}
Nicholas {Kaiser}, Herve {Aussel}, Barry~E. {Burke}, Hans {Boesgaard}, Ken
  {Chambers}, Mark~R. {Chun}, James~N. {Heasley}, Klaus-Werner {Hodapp}, Bobby
  {Hunt}, Robert {Jedicke}, D.~{Jewitt}, Rolf {Kudritzki}, Gerard~A. {Luppino},
  Michael {Maberry}, Eugene {Magnier}, David~G. {Monet}, Peter~M. {Onaka},
  Andrew~J. {Pickles}, Pui Hin~H. {Rhoads}, Theodore {Simon}, Alexander
  {Szalay}, Istvan {Szapudi}, David~J. {Tholen}, John~L. {Tonry}, Mark
  {Waterson}, and John {Wick}.
\newblock {Pan-STARRS: A Large Synoptic Survey Telescope Array}.
\newblock In J.~Anthony {Tyson} and Sidney {Wolff}, editors, {\em Survey and
  Other Telescope Technologies and Discoveries}, volume 4836 of {\em Society of
  Photo-Optical Instrumentation Engineers (SPIE) Conference Series}, pages
  154--164, December 2002.

\bibitem{gal2015dropout}
Yarin Gal and Zoubin Ghahramani.
\newblock Dropout as a bayesian approximation: Insights and applications.
\newblock {\em Deep Learning Workshop, ICML}, 1:2, 2015.

\bibitem{astronnleung2019deep}
Henry~W Leung and Jo~Bovy.
\newblock Deep learning of multi-element abundances from high-resolution
  spectroscopic data.
\newblock {\em Monthly Notices of the Royal Astronomical Society},
  483(3):3255--3277, 2019.

\bibitem{astronn}
Henry~W Leung.
\newblock Deep learning for astronomers with tensorflow.
\newblock {\em \href{GitHub repository}{https://github.com/henrysky/astroNN}.},
  2024. Accesed on 22/02/2024.

\bibitem{benatan2023enhancing}
Matt Benatan, Jochem Gietema, and Marian Schneider.
\newblock {\em Enhancing Deep Learning with Bayesian Inference: Create more
  powerful, robust deep learning systems with Bayesian deep learning in
  Python}.
\newblock Packt Publishing, 2023.

\bibitem{DEAP_JMLR2012}
F\'elix-Antoine Fortin, Fran\c{c}ois-Michel {De Rainville}, Marc-Andr\'e
  Gardner, Marc Parizeau, and Christian Gagn\'e.
\newblock {DEAP}: Evolutionary algorithms made easy.
\newblock {\em Journal of Machine Learning Research}, 13:2171--2175, 07 2012.

\bibitem{de2014deap}
Fran{\c{c}}ois-Michel De~Rainville, F{\'e}lix-Antoine Fortin, Marc-Andr{\'e}
  Gardner, Marc Parizeau, and Christian Gagn{\'e}.
\newblock Deap: enabling nimbler evolutions.
\newblock {\em ACM SIGEVOlution}, 6(2):17--26, 2014.

\bibitem{speagle2020dynesty}
Joshua~S Speagle.
\newblock dynesty: a dynamic nested sampling package for estimating bayesian
  posteriors and evidences.
\newblock {\em Monthly Notices of the Royal Astronomical Society},
  493(3):3132--3158, 2020.

\bibitem{aubourg2015}
{\'E}ric Aubourg, Stephen Bailey, Julian~E Bautista, Florian Beutler, Vaishali
  Bhardwaj, Dmitry Bizyaev, Michael Blanton, Michael Blomqvist, Adam~S Bolton,
  Jo~Bovy, et~al.
\newblock Cosmological implications of baryon acoustic oscillation
  measurements.
\newblock {\em Physical Review D}, 92(12):123516, 2015.

\bibitem{simplemc}
JA~Vazquez, I~Gomez-Vargas, and A~Slosar.
\newblock Updated version of a simple mcmc code for cosmological parameter
  estimation where only expansion history matters.
\newblock {\em \href{GitHub repository}{https://igomezv.github.io/SimpleMC}.},
  2023. Accesed on 22/02/2024.

\bibitem{efstathiou2008limitations}
G~Efstathiou.
\newblock Limitations of bayesian evidence applied to cosmology.
\newblock {\em Monthly Notices of the Royal Astronomical Society},
  388(3):1314--1320, 2008.

\bibitem{nesseris2004comparison}
Savas Nesseris and Leandros Perivolaropoulos.
\newblock Comparison of cosmological models using recent supernova data.
\newblock {\em Physical Review D}, 70(4):043531, 2004.

\bibitem{koo2022bayesian}
Hanwool Koo, Ryan~E Keeley, Arman Shafieloo, and Benjamin L'Huillier.
\newblock Bayesian vs frequentist: comparing bayesian model selection with a
  frequentist approach using the iterative smoothing method.
\newblock {\em Journal of Cosmology and Astroparticle Physics}, 2022(03):047,
  2022.

\bibitem{vazquez2012bayesian}
J~Alberto Vazquez, AN~Lasenby, M~Bridges, and MP~Hobson.
\newblock A bayesian study of the primordial power spectrum from a novel closed
  universe model.
\newblock {\em Monthly Notices of the Royal Astronomical Society},
  422(3):1948--1956, 2012.

\bibitem{hee2016bayesian}
Sonke Hee, WJ~Handley, Mike~P Hobson, and Anthony~N Lasenby.
\newblock Bayesian model selection without evidences: application to the dark
  energy equation-of-state.
\newblock {\em Monthly Notices of the Royal Astronomical Society},
  455(3):2461--2473, 2016.

\bibitem{handleyfgivenx}
Will Handley.
\newblock fgivenx: Functional posterior plotter.
\newblock {\em J. Open Source Softw.}, 3(28), Aug 2018.

\bibitem{snedecor1989statistical}
George~W. Snedecor and William~G. Cochran.
\newblock {\em Statistical Methods}.
\newblock Iowa State University Press, Ames, Iowa, 8 edition, 1989.

\end{thebibliography}
\end{document}